\documentclass[fleqn,usenatbib]{mnras}  
\usepackage{newtxtext,newtxmath}
\usepackage{xfrac}

\usepackage{ae,aecompl}

\usepackage{graphicx}	
\usepackage{amsmath}	
\usepackage[dvipsnames]{xcolor} 
\usepackage{ulem}
\usepackage{microtype}
\usepackage[outline]{contour}
\usepackage{tikz}

\newcommand{\be}{\begin{equation}}
\newcommand{\ee}{\end{equation}}
\newcommand{\fett}[1]{\boldsymbol{#1}}
\newcommand{\dd}{{\rm d}}
\newcommand{\ii}{{\rm{i}}}

\newcommand{\RT}{{\mathfrak{R}_{D}}}
\newcommand{\fb}{f_{\rm b}}
\newcommand{\fc}{f_{\rm c}}
\newcommand{\rb}{{\rm b}}
\newcommand{\rc}{{\rm c}}
\newcommand{\rM}{{\rm m}}
\newcommand{\bc}{{\rm bc}}
\newcommand{\rL}{{\rm L}}
\newcommand{\ini}{{\rm ini}}

\newcommand{\twoPPT}{\text{\scriptsize 2PPT}}
\newcommand{\free}{\text{\scriptsize free}}
\newcommand{\fixedxptp}{\text{\scriptsize $\fett{x}', D'$}}
\newcommand{\nab}{\fett{\nabla}}

\definecolor{lime}{HTML}{A6CE39}
\DeclareRobustCommand{\orcidicon}{
	\begin{tikzpicture}
	\draw[lime, fill=lime] (0,0) 
	circle [radius=0.14] 
	node[white] {{\fontfamily{qag}\selectfont \tiny ID}};
	\draw[white, fill=white] (-0.0625,0.095) 
	circle [radius=0.007];
	\end{tikzpicture}
	\hspace{-2mm}
}

\foreach \x in {A, ..., Z}{%
	\expandafter\xdef\csname orcid\x\endcsname{\noexpand\href{https://orcid.org/\csname orcidauthor\x\endcsname}{\noexpand\orcidicon}}
}

\contourlength{0.3pt} 
\contournumber{10} 

\title[Two-fluid perturbation theory]{Cosmological perturbations for two cold fluids in \contour{black}{$\Lambda$}CDM}

\author[Rampf et al.]{
 Cornelius Rampf$^{\,{\tiny\orcidA{}}\,\,\,\,\hyperlink{OCA}{1}}$\thanks{\!\!Marie Sk\l odowska--Curie Fellow; e-mail: \href{mailto:cornelius.rampf@oca.eu}{cornelius.rampf@oca.eu}}, Cora Uhlemann$^{\,{\tiny \orcidB{}}\,\,\,\,\hyperlink{NUC}{2}}$, and Oliver Hahn$^{\,{\tiny\orcidC{}}\,\,\,\,\hyperlink{OCA}{1}}$  \hypertarget{OCA} \\
$^{1}$Universit\'e C\^ote d'Azur, Observatoire de la C\^ote d'Azur, CNRS, Laboratoire Lagrange, \\\quad\!Boulevard de l'Observatoire, CS 34229, 06304 Nice, France \hypertarget{NUC}\\
$^{2}$School of Mathematics, Statistics and Physics, Herschel Building, Newcastle University,  \\\quad\!\!Newcastle upon Tyne, NE1 7RU, UK}

\date{Accepted XXX. Received YYY; in original form ZZZ}

\pubyear{2020}

\begin{document}
\label{firstpage}
\pagerange{\pageref{firstpage}--\pageref{lastpage}}
\maketitle

\begin{abstract}
The cosmic large-scale structure of our Universe is comprised of baryons and cold dark matter (CDM). 
Yet it is customary to treat these two components as a combined single-matter fluid with vanishing pressure, which is justified only for sufficiently large scales and late times. Here we go beyond the single-fluid approximation and develop the perturbation theory for two gravitationally coupled fluids while still assuming vanishing pressure. We mostly focus on perturbative expansions in powers of $D$ (or $D_+$), the linear structure growth of matter in a $\Lambda$CDM Universe with cosmological constant $\Lambda$. We derive in particular (1)~explicit recursion relations for the two fluid densities, (2)~complementary all-order results in the Lagrangian-coordinates approach, as well as (3)~the associated component wavefunctions in a semi-classical approach to cosmic large-scale structure. In our companion paper we apply these new theoretical results to generate novel higher-order initial conditions for cosmological hydrodynamical simulations. 
\end{abstract}

\begin{keywords}
 cosmology: theory -- large scale structure of Universe -- dark matter
\end{keywords}

\section{Introduction}

Analytical models for predicting the cosmic large-scale structure (LSS) are indispensable for interpreting cosmological observations, especially at high redshifts where cosmological perturbation theory (PT) is meaningful. 
In particular, accurate theoretical modelling is needed to extract and interpret cosmological data from the baryonic acoustic oscillation features imprinted in the statistics of the LSS \citep{2011MNRAS.416.3017B,2011MNRAS.418.1707B,2013AJ....145...10D,Slepian:2015,2018MNRAS.474.2109S}, or from tracers of the intergalactic medium which can be probed through absorption lines in the Lyman-$\alpha$ forest \citep{2006ApJS..163...80M,2010Natur.466..463C}.
Furthermore, having accurate PT predictions at hand is essential to reduce theoretical uncertaintities in the initial conditions for cosmological simulations; see e.g.\ \cite{Crocce:2006,Garrison:2016,Michaux:2020}.

According to the standard model of cosmology, dubbed $\Lambda$CDM, our Universe is comprised of cold dark matter (CDM), baryons, dark energy and relativistic species (radiation and neutrinos). After an initial inflationary phase of accelerated expansion, these species are  effectively coupled to each other through gravito-electroweak interactions. In full generality, the evolution of the relativistic and non-relativistic species is governed by the Einstein--Boltzmann equations. Predicting the LSS thus amounts to solving for these highly non-linear equations, which at this stage is not feasible. Instead it is customary to dissect the full problem into individual sub-problems and solve them for given temporal and spatial scales. For this analytical insight is of utmost importance, especially considering that some of the sub-problems are not decoupled, as we elucidate briefly in the following.

Dark energy is believed to be described by a cosmological constant $\Lambda$ -- which has no spatial dependence, and thus affects the evolution of the matter species through the global expansion of the Universe (parametrized by the usual Friedmann equations). The peculiar motion of matter, superimposed on the global expansion, is however not decoupled from the Friedmann equations and manifest, for example, as the Hubble drag in the law of momentum conservation.

Also radiation -- photons and massless neutrinos, affect the matter evolution. Before decoupling, radiation has a pre-dominant effect, especially through Compton scattering with baryons. As the Universe expands, Compton scattering becomes ineffective and radiation largely decouples from the peculiar matter evolution. This is because the mean density of radiation decays faster than the one of matter as the Universe expands, such that the impact of radiation on matter becomes less significant at late times. This argument can be demonstrated rigorously: Indeed, within the weak-field limit of general relativity and by employing tailor-made coordinate transformations, \cite{Fidler:2017pnb} has shown that the non-linear general relativistic equations of motion for matter can be brought precisely into the form of the Newtonian equations, {\it a.k.a.}\ the Euler--Poisson equations, which do not possess any couplings to radiation fluctuations. Their analysis reveals that any residual coupling between radiation and matter can be incorporated into a coordinate transformation, implying that a Newtonian theory (or simulation) for the matter evolution is meaningful on the considered scales. Additionally, the approach of \cite{Fidler:2017pnb} provides explicit instructions how the residual couplings can be efficiently incorporated {\it a posteriori,} i.e., after the Newtonian evolution was solved for (e.g., through an $N$-body simulation). Surprisingly, similar simplifications apply also when massive neutrinos are included, which has been very recently demonstrated by~\cite{Partmann2020}. Nonetheless we remark that there are recent attempts for incorporating the effects of massive neutrinos on matter in an active manner, see e.g.\ the numerical approaches of \cite{2017MNRAS.466L..68B,2019JCAP...03..022T} or the analytical approaches of~\cite{2014JCAP...11..039B,2020arXiv200706508A}.

Thus, determining the LSS can be effectively reduced to solving the Newtonian equations of baryons and CDM.
The problem focused originally on solving not for the individual baryons and CDM but for a combined, single-matter fluid, governed by the Euler--Poisson equations for sufficiently early times; see e.g.\ \cite{Bernardeau:2002} for a review. These single-fluid equations can be solved by using PT, either in Eulerian or Lagrangian coordinates. For the former, limitations of Eulerian PT have been known for quite a while; see e.g.\ \cite{2014JCAP...01..010B,2014PhRvD..89b3502B,2016PhLB..762..247N} for contemporary discussions. In essence, Eulerian PT does not predict observables, such as the matter power spectrum or bispectrum, to sufficient accuracy.  This lack in accuracy can be mostly explained by noting that standard Eulerian PT is fairly inefficient for resolving convective fluid motion and furthermore can not incorporate shell-crossing, which is the instance when the particle trajectories cross the first time \citep[see e.g.][]{Pueblas:2009}. As a consequence, in recent years many extensions or variants of Eulerian PT have been developed to circumvent this shortcoming; see e.g.\ the renormalized PT by \cite{Crocce:2006}, the time-flow renormalization approach of \cite{Pietroni:2008jx}, or the semi-numerical approaches of \cite{2012JCAP...01..019P,2014JCAP...09..047M}  and \cite{2017arXiv170704698S,Porto:2013qua,2015JCAP...10..039A,2020JCAP...03..018L} that marry extensions of PT with numerical (or observable) input.

Similar avenues for the single fluid have been pursued in Lagrangian-coordinates approaches, first by developing the foundations of Lagrangian PT \citep[see e.g.][]{Zeldovich:1970,Buchert:1989xx,Buchert:1992ya,1992ApJ...394L...5B,Bouchet:1994xp,Ehlers:1996wg,Zheligovsky:2014,Vlah:2014nta}, and subsequently by developing extensions thereof; see e.g.\ \cite{Matsubara:2007wj,2012JCAP...06..018R} for Lagrangian resummation schemes, the convolution Lagrangian PT approach of \cite{Carlson:2012bu}, or the Lagrangian effective field theory of LSS by \cite{Porto:2013qua}. While Lagrangian-coordinates approaches are very efficient for resolving convective motion, the standard approach still breaks down at shell-crossing. We however remark that there recent attempts for resolving the shell-crossing and post-shell-crossing regime on a deterministic level; see respectively \cite{Saga:2018,Rampf:2017jan,Rampf:2017tne} and \cite{Colombi:2014lda,Taruya:2017,Rampf:2019nvl,2020arXiv200211357C}.

Yet, none of these works treat baryons and CDM separately, and the physical motivation behind this simplification is that at sufficiently late times, baryons should follow the gravitational footprints made by CDM. While this appears to be a good approximation on large scales and late times, it is clear that incorporating relative effects between CDM and baryons is the next refinement step on the theory side; see e.g.\ \cite{2005MNRAS.362.1047N,Tseliakhovich:2010,2015JCAP...05..019L,Schmidt:2016,2020arXiv200709484G}. Highly related to the present paper are the approaches of~\cite{2009ApJ...700..705S,Somogyi:2009mh,2012PhRvD..85f3509B}. 

Specifically, \cite{2009ApJ...700..705S} introduced a novel approach for two fluids in PT up to third order including baryonic pressure \citep[see also][]{Chen:2019cfu}. While baryonic pressure is certainly highly relevant close to the Jeans scale, in the present paper we focus on rather large scales where its impact should be small. Our approach is possibly closer to the one of~\cite{Somogyi:2009mh,2012PhRvD..85f3509B} who also work in the limit of vanishing pressure, thereby assuming effectively two separate sets of identical fluid equations for CDM and baryons that are connected via a shared gravitational potential. Furthermore, while~\cite{Somogyi:2009mh} developed a renormalized PT for the two-fluid setup -- and~\cite{2012PhRvD..85f3509B} the eikonal approximation, which could potentially also model some shell-crossing effects, we are here focusing on times when shell-crossing dynamics are not yet dominant.  One of our motivations is to develop the necessary tools to provide accurate initial conditions for two-fluid cosmologies, both in Eulerian and Lagrangian coordinates which we directly exploit and compare against similar numerical avenues \citep[e.g.][]{Angulo:2013,Valkenburg:2017,Bird:2020} in our companion paper~\citep{HahnAccomp2020}.

In this paper we develop various PT approaches for the cold two-fluid model. The general methodology aims to take all decaying modes into account. Nonetheless, one of our main focuses here is to exploit certain boundary conditions that select, in the two-fluid case, the fastest growing modes {\it as well as a constant mode in the difference of the linear fluid densities,} which amounts to including the leading-order effects in the considered two-fluid model. For these growing-mode solutions, we are actually  able to provide explicit recursion relations to all orders. In addition to the classical approaches in Eulerian and Lagrangian coordinates, we also extend here the semi-classical description of \cite{Uhlemann:2019}, called Propagator Perturbation Theory (PPT), by generalizing their findings to a $\Lambda$CDM cosmology and to allow for two coupled fluids. Apart from circumventing some of the shortcomings of Eulerian PT, which is in particular resolving the inaccuracies of modelling convective motion (see Section~\ref{sec:schroedi} for further arguments), PPT is particularly suited  for  Ly-$\alpha$ studies \citep{Porqueres:2020} as well as generating initial conditions for simulations that require Eulerian fields as input.

It is worth noting here an appropriate physical picture for our model. As we elucidate in detail later on, the common gravitational potential of the baryonic and CDM components is sourced by the sum of their weighted densities which, by definition, is the density of a total matter fluid. Now, if that single-matter source in the gravitational potential is described in terms of purely growing-mode solutions (as it is very common in the literature), then it becomes evident that the individual fluid motion of the baryonic and CDM components must be identical in the growing mode, simply as a consequence of Newton's second law of motion. The component densities, however and crucially, generally differ, due to employing rigorously the boundary conditions that come with growing-mode solutions. Thus, in some sense, the employed approach for the growing-mode boils down to propagating initial density fluctuations along the paths of the respective fluids.

This paper is structured as follows. In \S\ref{sec:singleEuler} we begin with the Eulerian-coordinates approach for a single matter fluids, and explain the appropriate boundary conditions for selecting growing-mode solutions, which also are crucial for avenues beyond single fluid models. \S\ref{sec:2fluidEuler} generalizes the approach to two shared fluid components, where we report explicit all-order recursion relations for the difference of the fluid densities arising from non-decaying modes. In \S\ref{sec:singlefluidLPT} we review the Lagrangian-coordinates approach for a single fluid, while we generalize to two fluids in \S\ref{sec:2fluidLag}. A variational approach to single and two fluids, which employs contact geometry (an extension of symplectic geometry) is introduced in \S\ref{sec:H&HJE}, which largely serves as a classical analogue of the semi-classical description that we discuss in \S\ref{sec:schroedi}. Finally, we summarize our results and provide an outlook in~\ref{sec:conclusion}.

\paragraph*{Notation and nomenclature.} Eulerian coordinates are denoted with $\fett{x}$, while $\fett{q}$ refer to Lagrangian (or initial) coordinates. We use italic Latin letters for referring to spatial indices, summation over repeated indices is implicitly assumed, and we denote partial derivatives with the comma notation, i.e., $\nabla_{i} F = F_{,i}$. When there is risk of confusion, we reserve the comma notation for Lagrangian derivatives and the slash notation for Eulerian space derivatives, i.e.,  $\nabla_{x_i} G = G_{|i}$. As regards to temporal derivatives, we use the overdot for denoting the Lagrangian (or convective) time derivative with respect to the linear growth time~$D$. When a single fluid is considered, we attach an ``m'' to the fields, while in the two-fluid case the individual fluids have the roman labels ``b'' and ``c'', which are occasionally summarized with a (non-running!) Greek label~$\alpha = \rb, \rc$.

\section{Single fluid in Eulerian coordinates} \label{sec:singleEuler}

We begin by introducing the basic equations for a single cosmological fluid with vanishing pressure in the Newtonian limit, which are usually called Euler--Poisson equations. Throughout this paper we assume a $\Lambda$CDM Universe. After formulating the Euler--Poisson equations in suitable coordinates in \S\ref{sec:basicsingle}, for which we  employ the linear  growth factor $D$ as the time variable, we discuss the used boundary conditions in \S\ref{sec:slaving} that justify power expansions around $D=0$. Alternative perturbative expansions schemes are common in the literature and discussed in \S\ref{sec:alt-exp}.

\subsection{Basic equations} \label{sec:basicsingle}

We define the peculiar velocity with $\tilde{\fett{u}}_\rM = \partial_t \fett{x}$ where $\fett{x} = \fett{r}/a$ are the usual comoving coordinates and $a$ the cosmic scale factor which grows as~$t^{2/3}$ for small $t$. We surmount dependent variables with a tilde when the cosmic time $t$ is used as an independent time variable. Further we define the matter density contrast $\tilde \delta_\rM = (\tilde \rho - \tilde{\bar \rho}(t))/\tilde{\bar \rho}(t)$ where $\bar \rho(t) \sim a^{-3}$. In these variables, the Eulerian fluid equations for a $\Lambda$CDM Universe are \citep[cf.][]{Rampf:2015mza}
\begin{subequations} \label{eq:fluidcosmic}
 \begin{align}
& \partial_t \tilde{\fett{u}}_\rM +  \tilde{\fett{u}}_\rM \cdot \nab \tilde{\fett{u}}_\rM = - 2 H \tilde{\fett{u}}_\rM  - \frac{1}{a^2} \nab \tilde \varphi \,,   \label{eq:Euler}  \\
& \partial_t \tilde\delta_\rM +\nab \cdot \left[ (1+\tilde\delta_\rM) \tilde{\fett{u}}_\rM\right] =0 \,, \label{eq:mass} \\
&\nab^2  \tilde{\varphi} = \frac{3}{2a} \tilde{\delta}_\rM \,,  \label{eq:Pois}
 \end{align}
\end{subequations}
where $H = (\partial_t a)/a$ is the Hubble parameter governed by the (first) Friedmann equation, defined here with $H^2 =1/a^3 + \Lambda$ where $\Lambda = \Omega_\Lambda/\Omega_\rM$. Linearizing the fluid equations, one obtains the standard ODE for the linear density fluctuations
\be \label{eq:evoL}
  \partial_t^2{\tilde{\delta}}_\rM  +2 H \partial_t{\tilde\delta}_\rM = \frac{3}{2a^3} \tilde\delta_\rM \,.
\ee
Its solution is most easily obtained by changing from cosmic time to $a$-time: the growing-mode solution is 
\be \label{eq:Dplus}
  D(a) = a \sqrt{1+ \Lambda a^3} \, {}_2F_1\!\left( 3/2, 5/6, {11}/{6}, - \Lambda a^3 \right)\,,
\ee 
where ${}_2F_1$ is the Gauss hypergeometric function, while the other solution is decaying as $\sqrt{1+ \Lambda a^3} a^{-3/2}$~\citep[see e.g.][]{Demianski:2005is}.

Analytic solutions for arbitrary short times are only feasible when growing-mode solutions are selected. Indeed, the decaying solution blows up for $a \to 0$ invalidating linearization, while the growing-mode solution is analytic and has the small-$a$ expansion $D(a) = a- (2/11)\Lambda a^3 + O(a^7)$.  When we later seek perturbative expansions in powers of the growing-mode~$D$, it  will turn out to be essential to change the temporal dependence in the fluid equations to the growing mode~$D$. Defining a new velocity variable $\fett{v}_\rM \equiv \partial_D \fett{x} =  \tilde{\fett{u}}_\rM/\partial_t D$ and setting $\tilde \delta_\rM(t) = \delta_{\rm m}(D)$, we can recast the fluid equations to
\begin{subequations} \label{eqs:fluidD}
 \begin{align}
& \partial_D \fett{v}_\rM + \fett{v}_\rM \cdot \nab \fett{v}_\rM = - \frac{3 g}{2D} \big( \fett{v}_\rM + \nab \varphi \big) \,,   \label{eq:EulerD}  \\
& \partial_D \delta_\rM +\nab \cdot \big[ (1+\delta_{\rm m}) \,\fett{v}_\rM \big] =0 \,, \label{eq:massD} \\
&\nab^2  \varphi = \frac {\delta_{\rm m}} D \,,  \label{eq:PoisD}
 \end{align}
\end{subequations}
where $\tilde \varphi = 3D \varphi/(2a)$, and we have defined 
\be \label{eq:defg}
  g = g(D) = (D/\partial_t D)^2 a^{-3} = 1 + \Lambda D^3/11 + O(D^6)\,,
\ee 
which is analytic for small $D$ (and for small $a$, too). 
Thus, $g\approx 1$ up to third order in PT (see also Fig.\,\ref{fig:DandE}), justifying the validity of the following approximation of the ODE 
\be \label{eq:ODE-D}
  \partial_D^2 \delta_\rM + \frac{3}{2D} \partial_D \delta_\rM -  \frac{3}{2D^2}  \delta_\rM  =0\,,
\ee 
which has the general solution
\be
  \delta_\rM =  D \,C_+^\rM + D^{-3/2} C_-^\rM \,.
\ee
Here $C_+^\rM$ and $C_-^\rM$ are spatial integration constants for the standard growing and decaying solutions of linear density fluctuations, which can be fixed by providing suitable boundary conditions to~\eqref{eq:ODE-D} at sufficiently early times $D_{\rm ini}$. \cite{Buchert:1992ya} has shown that Zel'dovich-like solutions can be achieved with two types of boundary conditions, that either achieve $\delta_\rM(D_{\rm ini}) =0$ exactly, or to a very good approximation, assuming initial quasi-homogeneity;  see also \S3 of~\cite{Rampf:2012xa} for further details. Henceforth when discussing solutions including decaying modes we shall make use of the common assumption $\delta_\rM(D_{\rm ini}) =0$. This setting is actually essential for growing-mode solutions, as we elucidate in the following.

\subsection{Growing-mode solutions and slaving}\label{sec:slaving}

Observe that Eqs.\,\eqref{eqs:fluidD} are analytic for $D \to 0$ provided we impose the {\it slaved boundary conditions} \citep{Brenier:2003xs}
\be  \label{eq:slaving}
  \boxed{\delta_\rM^{\rm ini} = 0 \,, \qquad \qquad \fett{v}_\rM^{\rm ini} = - \nab \varphi^{\rm ini} \,,}
\ee
where ``ini'' denotes evaluation at $D=0$. As argued by \cite{Rampf:2015mza}, these boundary conditions imply initial quasi-homogeneity with zero vorticity ($\nab \times \fett{v}=0$). Furthermore, they provide the mathematical foundation for growing-mode solutions of the form
\be \label{eq:SPTmAnsatz}
  \delta_\rM = \sum_{n=1}^\infty \delta_\rM^{(n)}(\fett{x})\,D^n \,, \qquad  \theta_\rM = \nab \cdot \fett{v}_\rM = - \sum_{n=1}^\infty \theta_\rM^{(n)}(\fett{x})\,D^{n-1} \,,
\ee
where $\delta_\rM^{(n)}$ and $\theta_\rM^{(n)}$ are spatial coefficients that can be easily determined. For reference, the first- and second-order solutions are
\begin{align} \label{eq:2SPT}
 \begin{aligned}
   \delta_\rM^{(1)}\! &= \nab^2\! \varphi^{\rm ini} \,, \, \quad \delta_\rM^{(2)} \!= \frac 5 7 \varphi_{,ll}^{\rm ini} \varphi_{,mm}^{\rm ini} \!+ \varphi_{,llm}^{\rm ini} \varphi_{,m}^{\rm ini} + \frac 2 7 \varphi_{,lm}^{\rm ini} \varphi_{,lm}^{\rm ini} \,, \\
 \theta_\rM^{(1)} \! &= \nab^2 \!\varphi^{\rm ini} \,, \,\quad \theta_\rM^{(2)} = \! \frac 3 7 \varphi_{,ll}^{\rm ini} \varphi_{,mm}^{\rm ini} \!+ \varphi_{,llm}^{\rm ini} \varphi_{,m}^{\rm ini} + \frac 4 7 \varphi_{,lm}^{\rm ini} \varphi_{,lm}^{\rm ini} \,,
 \end{aligned}
\end{align}
where ``${,l}$'' denotes partial differentiation with respect to component $x_l$, and summation over repeated indices is assumed. We again like to stress that in the present expansion scheme, assuming $g=1$ is exact up to fourth order in PT.  Of course, other expansions may be employed, which we discuss next.

\subsection{Alternative expansions in \texorpdfstring{$\fett{\Lambda}$}{L}CDM}\label{sec:alt-exp}

While the perturbative solutions~\eqref{eq:2SPT} in $\Lambda$CDM are well-known in the literature \cite[e.g.][]{Bernardeau:2002}, their derivation is usually not based on a strict $D$ expansion. Instead it is customary to either employ fitting functions, or to derive the results in the Einstein--de Sitter (EdS) approximation and replace the respective growth functions according to $a \to D$,  which yields a fairly accurate approximation to the $\Lambda$CDM equations~\cite[see e.g.][]{Pietroni:2008jx,Hiramatsu:2009ki}.

\begin{figure}
	\centering
	\includegraphics[width=0.95\columnwidth]{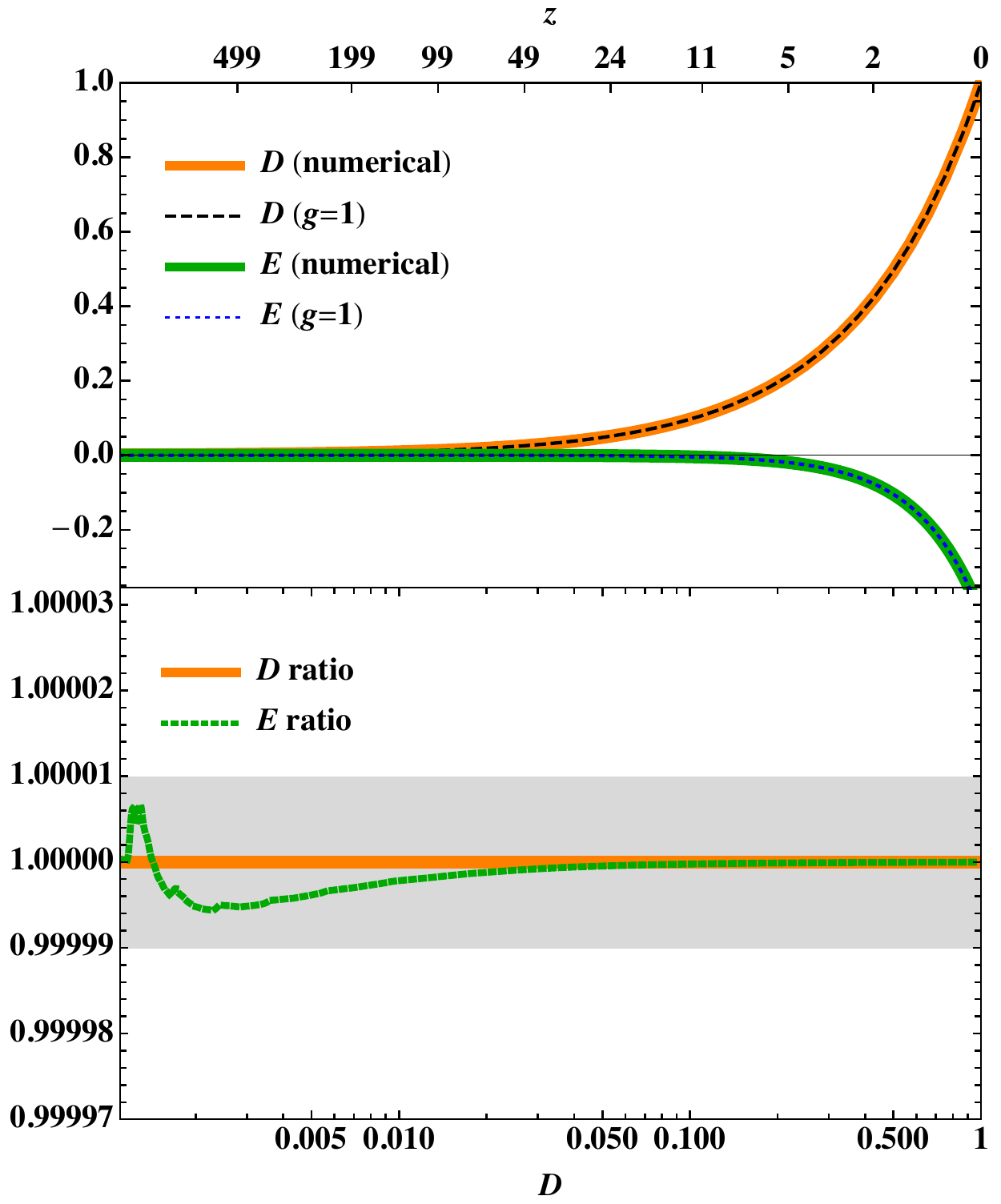}
	\caption{Temporal evolution of first- and second-order growth functions, denoted respectively with $D$ and $E$. {\bf Top:} Solid lines denote numerical solutions of the ODE's~\eqref{eq:evoL} and~\eqref{eq:ODE-E} where the full evolution of $g(D)$ is taking into account, while dashed and dotted lines denote the respective analytical approximation of these ODE's when $g=1$ for which $E =(-3/7)D^2$. {\bf Bottom:} Ratios of numerical solutions against their analytical approximation. The tiny deviation at earlier times for $E$ is of numerical nature, due to an evaluation of ratios where both the nominator and denominator tend to zero. 
}
	\label{fig:DandE}
\end{figure}

Apart from a strict expansion in powers of the growing mode, one could also solve for the temporal coefficients at each order separately. In our language, this amounts to incorporating the late-time evolution of $g$ as it could be already relevant at low perturbative orders. Such avenues may become relevant particularly when decaying modes are taken into account. In this scenario, one may impose a perturbative expansion in the ``weak'' sense, i.e., to not fix a physical expansion parameter and instead assume a certain smallness in the fields
\be
  \delta_\rM = \epsilon \,\delta_\rM^{(1)} + \epsilon^2 \,\delta_\rM^{(2)}+ \ldots \,,
\ee
and likewise for the velocity (and displacement) field, where $\epsilon$ is a small perturbation parameter which may be set to unity, once the perturbative equations are determined. Calculational details are provided in Appendix~\ref{app:1fluid2LPT}, while the results for the densities are
\begin{subequations}
\begin{align}
  &\delta_\rM^{(1)} =  D \nab^2 \varphi^{\rm ini} \,, \\
  &\delta_\rM^{(2)} =  \frac{\text{\small $D^2-E$}}{\text{\small $2$}} \varphi_{,ll}^{\rm ini} \varphi_{,mm}^{\rm ini} \!+  \text{\small $D^2$} \varphi_{,llm}^{\rm ini} \varphi_{,m}^{\rm ini} + \frac{\text{\small $D^2 +E$}}{\text{\small $2$}} \varphi_{,lm}^{\rm ini} \varphi_{,lm}^{\rm ini} \,, \label{eq:soldelta2weak}
\end{align}
\end{subequations}
where $E$ is a second-order temporal coefficient, subject to the ODE
\be \label{eq:ODE-E}
  \partial_D^2 E + \frac{3g}{2D} \partial_D E - \frac{3g}{2D^2} E = - \frac{3g}{2}\,.
\ee
Evidently, in the case when $g=1$, as effectively employed for growing-mode expansions,  the solution to this ODE can be analytically determined, with the fastest-growing mode $E \to -(3/7) D^2$.  
Numerical results to~\eqref{eq:ODE-E}  can be found in \cite{Bouchet:1994xp}, while analytical solutions involving hypergeometric functions are provided by~\cite{Matsubara:1995kq}.
In Fig.\,\ref{fig:DandE} we show that, for the two temporal coefficients involved, the analytical solutions for $g=1$, against their respective numerical solutions where $g$ is fully taken into account. The agreement between the analytical and  numerical solutions is excellent. We thus conclude that $g$ can be safely set to unity, at least for the present task. Similar conclusions have been drawn by~\cite{Tram:2016cpy} who also investigated numerically the solution of the ODE~\eqref{eq:ODE-E}; however, their numerical result departs at late times close to $a\simeq 1$ from the analytical prediction by a few percent (see their Fig.\,6), which we speculate may be of numerical~nature.

We note again that when (all) decaying modes are included in the analysis, which is not our main focus, the late-time evolution of~$g$ could eventually become important at successive higher orders.

\section{Two cold fluids in Eulerian coordinates}
\label{sec:2fluidEuler}

Let us now turn to the Eulerian formulation for two fluids. These two fluids can be thought of as modelling the individual evolution of baryons and CDM in our Universe, hence the labels ``b'' and ``c'' that we use throughout this paper. Nonetheless we remark that we do not consider the effects of baryonic pressure, which  limits our theoretical predictions to scales larger than the Jeans length. After introducing the relevant equations and linearized solutions including decaying modes in \S\ref{sec:2fgeneral}, we generalize the slaving conditions to the two-fluid case and derive explicit all-order recursion solution in \S\ref{sec:growing2SPT2F}. 
We remark that the present considerations can be easily generalized to more than two cold fluids; we shall come back to this in \S\ref{sec:conclusion}.

\subsection{General formalism} \label{sec:2fgeneral}

Consider two fluids $\alpha = \rb, \rc$ that are gravitationally coupled via
\begin{subequations} \label{eqs:fluidsD}
 \begin{align}
& \partial_D \fett{v}_\alpha + \fett{v}_\alpha \cdot \nab \fett{v}_\alpha = - \frac{3 g}{2D} \big( \fett{v}_\alpha + \nab \varphi \big) \,,   \label{eq:EulerDalpha}  \\
& \partial_D \delta_\alpha +\nab \cdot \big[ (1+\delta_\alpha) \,\fett{v}_\alpha \big] =0 \,, \label{eq:massDalpha} \\
&\nab^2  \varphi = \frac {\delta_\rM} D \,,  \label{eq:PoisDalpha}
 \end{align}
\end{subequations}
where 
\be
  \delta_\rM = \fb \delta_\rb + \fc \delta_\rc \,, \qquad \quad \fb + \fc = 1 \,,
\ee 
with the present (baryon) fraction $\fb = \Omega_\rb/\Omega_\rM$, with $\Omega_\rM = \Omega_\rb + \Omega_\rc$. To proceed, it is instructive to work with the sum and difference of the Euler equation~\eqref{eq:EulerDalpha} for the fluid components. 
A ``sum'' equation for~\eqref{eq:EulerDalpha} is obtained by first multiplying the equation for b by $\fb$ and c by $\fc$, and sum up those equations. Similarly, one proceeds with~\eqref{eq:massDalpha}. Linearizing the sum and difference equations we have 
\begin{subequations} \label{eqs:linearised2fluid}
\begin{align} 
  &\partial_D  \fett{v}_{\rm m} = -  \frac{3g}{2D} \left(  \fett{v}_{\rm m} + \nab \varphi \right) \,, \quad 
       &\partial_D \delta_{\rm m} + \nab \cdot \fett{v}_{\rm m} = 0 \,,  \label{eq:matterlinear} \\
 &\partial_D  \fett{v}_{\rm bc}  = - \frac{3g}{2D} \fett{v}_{\rm bc} \,,  
       &\partial_D \delta_{\rm bc} + \nab \cdot \fett{v}_{\rm bc} = 0 \,, \label{eq:differencelinear}  \\
  & \nab^2 \varphi = \frac {\delta_{\rm m}} D 
\end{align}
\end{subequations}
(see App.\,\ref{app:sumdifference-nonpert} for a non-perturbative generalization of these equations),
where $\delta_{\rm bc} = \delta_\rb - \delta_\rc$,  $\fett{v}_{\rm bc} = \fett{v}_\rb - \fett{v}_\rc$, and $\fett{v}_{\rm m} = \fb \fett{v}_\rb+ \fc \fett{v}_\rc$. The linearized equations for the sum and difference decouple \citep[cf.][]{Schmidt:2016}, leading to the separate evolution equations 
\be
 \begin{aligned}
  &\partial_D^2 \delta_\rM + \frac{3g}{2D} \partial_D \delta_\rM = \frac{3g}{2D^2}  \delta_\rM \,, \\ 
    &\partial_D^2 \delta_\bc + \frac{3g}{2D} \partial_D \delta_\bc =0 \,.
 \end{aligned} 
\ee
For $g =  1$ (justified in \S\ref{sec:alt-exp} and through Fig.\,\ref{fig:DandE}), the general analytical solutions are
\be
  \delta_\rM = D \,C_+^\rM + D^{-3/2} C_-^\rM \,, \qquad \quad  \delta_\bc =  \delta_\bc^\ini -2 D^{-1/2} C_-^\bc \,,
\ee
which likewise can be used to express the general solutions for the components at first order,
\begin{align} \label{db}
 \begin{aligned}
   \delta_\rb &= D \,C_+^\rM + D^{- 3/2}\, C_-^\rM   - 2 D^{-1/2}  \fc C_-^\bc  + \delta_\rb^{\rm ini}  \,, \\
   \delta_\rc &= D \, C_+^\rM + D^{-3/2}\, C_-^\rM   + 2 D^{-1/2}  \fb C_-^\bc   + \delta_\rc^{\rm ini} \,, 
 \end{aligned}
\end{align}
where the initial densities are given by $\delta_\rb^{\rm ini} = \fc \delta_\bc^\ini$ and $\delta_\rc^{\rm ini} = - \fb \delta_\bc^\ini$. As we shall show in the following, these initial densities can not be set to zero in general, especially not when growing-mode solutions are considered.

\subsection{Slaving and all-order growing-mode solutions}\label{sec:growing2SPT2F}

Similarly  to the single-fluid case, we can deduce from the fluid equations~\eqref{eqs:linearised2fluid} the necessary conditions that guarantee their regularity for $D \to 0$. Indeed, we find that {\it in the two-fluid case the slaved boundary conditions} at $D=0$ are
\begin{align} \label{eq:slaving2fluid}
 \boxed{   \fett{v}_{\rm bc}^{\rm ini} \to 0 \,, \qquad \delta_{\rm m}^{\rm ini} \to 0 \,, \qquad  \fett{v}_{\rm m}^{\rm ini} = - \nab \varphi^{\rm ini} \,, }
\end{align}
which, from~\eqref{db}, translate into ``growing-'' and ``constant/persisting modes'' for the two fluids; at first order they read
\begin{align} \label{eq:deltaalpha2fl}
  \delta_\rb = D \,\nab^2 \varphi^{\rm ini} + \delta_\rb^{\rm ini} \,, \qquad \delta_\rc = D \,\nab^2 \varphi^{\rm ini} + \delta_\rc^{\rm ini} \,,
\end{align}
where we have used that  $C_+^\rM = \nab^2 \varphi^{\rm ini}$ (cf.\,Eq.\,\eqref{eq:2SPT}).
It is crucial to note that in the presence of two (shared) fluids, 
the weighted sum $\delta_\rM = \fb \delta_\rb + \fc \delta_\rc$ must sum up to zero initially, which requires a compensating relationship between $\delta_\rb^\ini$ and $\delta_\rc^\ini$. Thus, in general the initial densities~$\delta_\alpha^\ini$ must be nonzero, of course except in special points where both densities are zero on their own.

We remark that the evaluation of the fields at $D=0$ {\it does  not imply} that we ignore the inflationary and recombination dynamics; instead we actually reduce all physics prior to recombination to an infinitely thin boundary layer. As is discussed in detail by e.g.\ \cite{Michaux:2020}, this procedure is actually implicitly assumed when initializing Newtonian simulations in the growing mode, which is the standard in numerical cosmology. In the literature, the respective approach for generating the initial fields is usually called ``backscaling''; for details and further remarks we kindly refer to \S2.5 in our companion paper.

Observe that growing-mode results, such as~\eqref{eq:deltaalpha2fl}, can be obtained in a more direct way, i.e., 
a way that does not require the introduction of the weighted sum and difference variables, but instead assumes that the gravitational force of the two shared fluid system is identical to the one in the (combined) single matter case.
For this let us linearize (again) Eqs.\,\eqref{eqs:fluidsD} but now express the Poisson source in~\eqref{eq:PoisDalpha}
by the fastest-growing solution of the single matter density that we have discussed in \S\ref{sec:slaving}.
At first order one easily obtains 
\be \label{eq:ODEdeltaalphaPoissonsource}
  \partial_D^2 \delta_\alpha + \frac{3}{2D} \partial_D \delta_\alpha = \frac{3}{2} \nab^2 \varphi^{\rm ini} 
\ee
for $g =1$,
which has the solution $\delta_\alpha = D \nab^2 \varphi^{\rm ini} - 2 C_1^\alpha D^{-1/2} + C_2^\alpha$. Clearly, in the case of slaved
boundary conditions which only select non-decaying terms, $C_1^\alpha$ must vanish. Similarly, by identification with the general solution~\eqref{db}, 
we have $C_2^\alpha =  \delta_\alpha^{\rm ini}$. Thus, we arrive at the identical result as above, without first solving separately for the sum and difference variables.

Physically, this procedure works as in the present case with growing-mode solutions, we are essentially just splitting up a single matter fluid
into two shared components. While the shared fluid components must begin their evolution with non-zero initial densities (due to slaving), the evolved 
matter density in the growing mode must be equal to the growing mode of the sum of the two shared component densities.

This discussed simplification carries over to all orders. For example, truncated to second order we find the following solutions for $\alpha = \rb, \rc$, 
\begin{align} \label{eq:2SPT2fluids}
\boxed{ \begin{aligned}
   &\!\delta_\alpha(\fett{x}, D)  = D \, \delta_\rM^{(1)} + \delta_\alpha^{\rm ini} +  D^2 \delta_\rM^{(2)} + D \!\left(  \delta_\alpha^{\rm ini} \varphi_{,ll}^{\rm ini}  +  \delta_{\alpha, l}^{\rm ini} \varphi_{,l}^{\rm ini}   \right) , \\
   &\!\theta_\alpha(\fett{x}, D) = \theta_\rM^{(1)} + D\, \theta_\rM^{(2)}  ,
 \end{aligned}
 }
\end{align}
where the growing-mode solutions for the total matter density and velocity are given in Eqs.\,\eqref{eq:2SPT}. 
Furthermore, imposing
\be 
  \delta_\bc = \sum_{n=1}^\infty \delta_\bc^{(n)}(\fett{x}) \,D^{n-1} \,,
\ee
we find simple all-order recursion relations for the difference density, 
\be \label{eq:recdeltabcMT}
 \boxed{  \delta_\bc^{(n)} = \frac{1}{n-1} \sum_{0<s<n} \nab \cdot \left[ \delta_\bc^{(s)}  \nab^{-2} \nab \theta_\rM^{(n-s)} \right] }
\ee
for $n>1$, and $\delta_\bc^{(1)}= \delta_\bc^\ini$ for $n=1$, where the coefficients $\theta_\rM^{(s)}$ are defined in Eq.\,\eqref{eq:SPTmAnsatz}.
See Appendix~\ref{app:2fluid2SPT} for calculational details, where we also provide explicit formulas to determine the power spectrum of $\delta_\bc$ to one-loop accuracy. Using the well-known recursion relations for the matter density and velocity from standard PT, as well as the recursion relation for the difference density, it is clear that the  $\delta_\alpha$'s can be easily determined to any desired level of accuracy, namely by iteratively solving for $\delta_\alpha$ through the two identities $\delta_\rM = \fb \delta_\rb + \fc \delta_\rc$ and $\delta_\bc \equiv \delta_\rb -\delta_\rc$, leading to 
\begin{align} \label{eq:solE2fluidbandc}
 \begin{aligned}
  &\delta_\rb \!=\!  \delta_\rM + \fc \delta_\bc  \!=\! \sum_{n=1}^\infty \left[ \delta_\rM^{(n)}(\fett{x}) \,D^{n} + \fc \delta_\bc^{(n)}(\fett{x})\, D^{n-1} \right] \,, \\
  &\delta_\rc \!=\! \delta_\rM  - \fb \delta_\bc  \!=\! \sum_{n=1}^\infty \left[ \delta_\rM^{(n)}(\fett{x}) \,D^{n} - \fb \delta_\bc^{(n)}(\fett{x})\, D^{n-1} \right] \,. 
  \end{aligned}
\end{align}
Low perturbative orders for $\delta_\rb$ and $\delta_\rc$ have been reported in the literature in various approaches \citep{2009ApJ...700..705S,Somogyi:2009mh,2015JCAP...05..019L}, however we are not aware of previous work on two-fluid solutions that include the all-order contributions from $\delta_\bc$ as instructed through~\eqref{eq:solE2fluidbandc};  it appears that this is due to the different nature of their approaches.
As previously mentioned, we argue  
that the constant mode~$\delta_\bc^\ini$ is non-negligible for purely growing-mode solutions in two-fluid cosmologies.

\section{Single fluid in Lagrangian coordinates}\label{sec:singlefluidLPT}

Lagrangian-coordinates approaches have a long tradition in cosmology, starting with pioneering works on fundamentals of \cite{Zeldovich:1970,Novikov:2010ta,Buchert:1989xx,1992ApJ...394L...5B,Ehlers:1996wg,Zheligovsky:2014,Rampf:2017jan,Saga:2018}, as well as a host of extensions thereof, such as \cite{2002PhRvD..66f4014T,Matsubara:2007wj,Carlson:2012bu,Porto:2013qua,Vlah:2014nta,2020arXiv200706508A} and many others.

Let $\fett{q} \mapsto \fett{x}(\fett{q},D) = \fett{q} + \fett{\xi}^\rM(\fett{q},D)$ be the Lagrangian map, from initial ($D=0$) position $\fett{q}$ to the current position $\fett{x}$ at time~$D$. In the Lagrangian representation, the velocity is defined by $\partial_D^{\rm L} \fett{x} = \dot {\fett{x}} = \fett{v}_\rM$, where $\partial_D^{\rm L}$  is the Lagrangian (convective) $D$-time derivative, which is here and in the following also denoted by an overdot. The Lagrangian time derivative commutes with the Lagrangian space derivative, however not with the Eulerian one. In the single fluid case, and before shell-crossing, mass conservation reads exactly 
\be 
 \delta_\rM(\fett{x}(\fett{q},D))  = \frac{1}{J} - 1 \,, 
\ee
 where $J$ is the Jacobian
\be
  J = \det[x_{i,j}] = 1 + \xi_{i,i}^\rM + \frac 1 2 \left[ \xi_{i,i}^\rM \xi_{j,j}^\rM- \xi_{i,j}^\rM \xi_{j,i}^\rM \right] + \det [\xi_{i,j}^\rM] \,,
\ee 
which, due to the choice of Lagrangian coordinates, is unity at initial time. To get Lagrangian evolution equations, one usually takes the Eulerian divergence of the Euler equation~\eqref{eq:EulerD} and considers the vanishing of the Eulerian vorticity, which respectively lead to
\begin{subequations} \label{eqs:fluidDLag}
 \begin{align}
& \nab_x \cdot \RT \fett{x}(\fett{q},D)  = - \frac{3 g}{2D^2} \delta_\rM(\fett{x}(\fett{q},D)) \,,   \label{eq:EulerDLag}  \\
& \nab_x \times \dot{\fett{x}}(\fett{q},D) =0 \,,  \label{eq:vorticity}
 \end{align}
\end{subequations}
where we have used the Poisson equation~\eqref{eq:PoisD} and defined the second-order temporal operator
\be \label{eq:Rop}
  \RT = (\partial_D^\rL)^2 +[3 g/(2D)] \partial_D^\rL\,.
\ee 
Equations~\eqref{eqs:fluidDLag} are not yet fully written in Lagrangian space since the density should be expressed in terms of the Jacobian; furthermore there are still remaining Eulerian derivatives. Regarding the latter, \cite{Buchert:1987xy} suggested to proceed as follows: convert the derivatives to Lagrangian space, i.e., $\nabla_{x_i} = (\partial q_j/\partial x_i) \nabla_{q_j}  = x_{i,j}^{-1} \nabla_{q_j}$, followed by representing the components of the matrix inverse of $x_{i,j}$ by employing the adjugate, i.e., $x_{i,j}^{-1} = {\rm adj}(x_{i,j})/J =  \varepsilon_{ikl} \varepsilon_{jmn} x_{k,m} x_{l,n}/(2J)$, where $\varepsilon$ is the fundamental antisymmetric (Levi-Civita) tensor. Using this, Eqs.\,\eqref{eqs:fluidDLag} respectively become the final Lagrangian evolution equations in the single-fluid case, 
\begin{subequations} \label{eqs:fluidDLagFinal}
 \begin{align}
&  \varepsilon_{ikl} \varepsilon_{jmn} x_{k,m} x_{l,n}\, \RT  x_{i,j}  =  \frac{3 g}{D^2} \left( J-1 \right) \,,  \label{eq:Lag1}   \\
& \varepsilon_{ijk} \dot x_{l,j}  x_{l,k} = 0 \,, \label{eq:Lag2}
 \end{align}
\end{subequations}
where $g$ and $\RT$ are respectively defined in equation~\eqref{eq:defg} and~\eqref{eq:Rop}. 
All indices in Eq.\,\eqref{eq:Lag1} are contracted, hence it is a scalar equation. By contrast, Eq.\,\eqref{eq:Lag2} is a vector equation which states the conservation of the {\it Eulerian zero-vorticity condition.} 
Of course, equation~\eqref{eq:Lag2} can be generalized to allow for initial vorticity \citep[cf.][]{Rampf:2016wom}, turning the vectorial equation into {\it invariants equations} that have first been derived by Cauchy in 1815 for incompressible Euler flow; see \cite{Zheligovsky:2014} for further details and historical notes. Nevertheless, we remark that initial vorticity is not compatible with purely growing-mode solution, as also pointed out by \cite{Matsubara:2015ipa}; the mathematical reasoning for this statement is founded by the slaving conditions~\eqref{eq:slaving}, that, of course, also apply in Lagrangian coordinates.

Growing-mode solutions are achieved by plugging the power-law {\it Ansatz} for the displacement 
\begin{subequations} \label{eq:singleLPT}
\be
  \fett{x}-\fett{q} = \fett{\xi}^\rM (\fett{q},D) = \sum_{n=1}^\infty \fett{\zeta}^{\rM (n)}(\fett{q})\,D^n
\ee 
into the evolution equations~\eqref{eqs:fluidDLagFinal}, leading to the solutions 
\begin{align} \label{eq:sol2LPTsingle}
   \begin{aligned}
      {\zeta}_{l,l}^{\rM (1)} = - \nab^2 \varphi^{\rm ini} \,, \,\qquad 
      {\zeta}_{l,l}^{\rM (2)} = - \frac{3}{14} \left( \varphi_{,ii}^{\rm ini} \varphi_{,jj}^{\rm ini}-\varphi_{,ij}^{\rm ini} \varphi_{,ij}^{\rm ini} \right) 
 \end{aligned}
\end{align}
\end{subequations}
at order $n=1,2$.  As in the Eulerian case, these results are well-known \citep[e.g.,][]{1992ApJ...394L...5B,Scoccimarro:1997,Matsubara:2015ipa}, however, to our knowledge, they have not been derived within a $D$-expansion but within a weak expansion.
In Appendix~\ref{app:1fluid2LPT} we also derive second-order solutions in Lagrangian perturbation theory (LPT) employing such a weak expansion scheme (cf.\ \S\ref{sec:alt-exp}), however, in the growing mode we find that those solutions agree perfectly with the ones reported above.

Finally, as is well known for growing-mode solutions, the displacement remains potential in Lagrangian coordinates up to second order, as can be verified by plugging the solution~\eqref{eq:sol2LPTsingle} into the Eulerian zero-vorticity condition~\eqref{eq:Lag2}. This potential character is lost at third order in the growing mode \cite[see e.g.][]{Buchert:1994}, leading to a transverse component in the displacement. By contrast, when decaying modes are included, a transverse displacement appears already at the second order; see \cite{Buchert:1993xz} for the result in the EdS limit, while the respective $\Lambda$CDM result is given in Eq.\,\eqref{eq:cauchysol} for $\alpha \to \rM$ and $C_{-}^{\rm bc} \to 0$.

\section{Two fluids in Lagrangian coordinates}
\label{sec:2fluidLag}

Now we switch to the Lagrangian-coordinates approach for two fluids. The governing equations are provided in \S\ref{sec:generalLag2F} and general perturbative solutions developed in \S\ref{sec:pertcoord}. In \S\ref{sec:altgrowingL2F} we provide a simplified derivation which is particularly suited for deriving all-order solutions in the growing modes. The solutions for the two-fluid displacements are particularly simple, however only when the initial densities are transported along the fluid paths. For numerical applications it may be instructive to absorb these initial densities in the displacements, which is discussed in \S\ref{sec:unpertcoord}.

\subsection{General formalism}\label{sec:generalLag2F}

Let $\fett{q} \mapsto \fett{x}^\alpha(\fett{q}, D) = \fett{q} + \fett{\xi}^\alpha (\fett{q},D)$ be the respective Lagrangian maps for the two fluids with corresponding displacements $\fett{\xi}^\alpha$, for $\alpha= {\rm b,c}$. Without loss of generality, we require that both component displacements vanish initially, which we will employ from here on, however except in \S\ref{sec:unpertcoord}, where we provide a complementary derivation where the component fluids are initialized in a ``perturbed'' Lagrangian coordinate system.

The component velocities are defined with $\fett{v}^\alpha = \dot {\fett{x}}^\alpha(\fett{q}, D)$. Using similar algebraic manipulations as in the previous section, we transfer the Eulerian two-fluid equations~\eqref{eqs:fluidsD} to Lagrangian space. We have firstly (see also \citealt{Chen:2019cfu})
\begin{subequations} \label{eq:L2F}
\begin{align}
  &\RT \fett{x}^\alpha = - \frac{3g}{2D} (\nab_x \varphi)_\alpha  \,, \label{Lag2fluid} \\
   &\delta_\alpha(\fett{x}^\alpha(\fett{q},D)) = (1+ \delta_\alpha^{\rm ini}(\fett{q}))/\det[\nab_q \fett{x}^\alpha(\fett{q},D)] -1 \,, \label{eq:massalphaLag} \\
   &(\nab^2_x \varphi)_\alpha  = \frac 1 D \left[ \fb\, \delta_{\rm b}(\fett{x}^\alpha(\fett{q},D)) +\fc \,\delta_{\rm c}(\fett{x}^\alpha(\fett{q},D))  \right] \,, 
\end{align} 
\end{subequations}
where $g= 1 + \Lambda D^3/11 + O(D^6)$ (see Eq.\,\eqref{eq:defg}), while the initial density $\delta_\alpha^{\rm ini}$ and the temporal operator $\RT$ are respectively given in Eqs.\,\eqref{db} and~\eqref{eq:Rop}. Furthermore,  we have defined the abbreviation $(\nab_x \varphi )_\alpha = \nab_{x^\alpha} \varphi(\fett{x}^\alpha(\fett{q},D))$, and, for notational simplicity, have suppressed some obvious dependencies in~\eqref{eq:L2F}. Note that we have now two Poisson equations, one for each Lagrangian map, they read \citep{FrischSobo:2015}
\begin{align}
\label{eq:deltamLag}
 \begin{aligned}
   &(\nab_x^2 \varphi)_\rb  
   = \frac 1 D \left[ \frac{\fb (1+ \delta_{\rm b}^{\rm ini}(\fett{q}))}{J^{\rm b}(\fett{q})}
   + \frac{ \fc (1+\delta_{\rm c}^{\rm ini}(\fett{q}))}{
     \left. J^{\rm c}(\tilde{\fett{q}})\right|_{\tilde{\fett{q}} = \fett{x}_\rc^{-1} \circ \,\fett{x}^{\rm b}(\fett{q})}}  
       -1 \right] \,, \\
 &(\nab_x^2 \varphi)_\rc  
 = \frac 1 D \left[ \frac{\fb (1+\delta_{\rm b}^{\rm ini}(\fett{q}))}{
 \left. J^{\rm b}(\tilde{\fett{q}}) \right|_{\tilde{\fett{q}} = \fett{x}_{\rm b}^{-1} \circ\,\fett{x}^{\rm c}(\fett{q})}
 }
 +   \frac{\fc(1+\delta_{\rm c}^{\rm ini}(\fett{q}))}{ J^{\rm c}(\fett{q})} - 1\right] \,,
 \end{aligned}
\end{align}
where $J^\alpha(\tilde{\fett{q}}) = \det[x_{i,j}^\alpha(\tilde{\fett{q}})]$. Furthermore we employ the composition
 $\fett{x}_\rc^{-1} \circ \,\fett{x}^\rb = \fett{x}_\rc^{-1}(\fett{x}^\rb)$, where $\fett{x}_\alpha^{-1} = \fett{q}^\alpha(\fett{x})$ is the inverse Lagrangian map defined such that $\fett{x}^\alpha (\fett{q}^\alpha (\fett{x})) = \fett{x}$.
Compositions such as  $\fett{x}_\rc^{-1} \circ \,\fett{x}^\rb$ have the purpose of 
pulling the field of particles of species~$\rc$ back to its initial value, and then forward it to the current time such that the Poisson equation of species~$\rb$ takes the gravitational interactions of~$\rc$ at the current position~$\fett{x}^\rb$ into account.

Finally, these equations can be combined leading to the Lagrangian fluid equations for the 2-fluid system,
\begin{subequations} \label{eq:mainL2F}
\begin{align} 
   & \varepsilon_{ikl} \varepsilon_{jmn}  x_{k,m}^\alpha  x_{l,n}^\alpha \RT x_{i,j}^\alpha = - \frac{3g}{D} J^\alpha (\nab_x^2 \varphi)_\alpha \,,  \label{eq:LEOM1} \\
  & \varepsilon_{ijk} x_{l,j}^\alpha \dot x_{l,k}^\alpha =0 \,,  \label{eq:LEOM2}
\end{align}
\end{subequations}
where we remind the reader that $\alpha = \rb, \rc$ are non-running fluid labels, while summation over Latin indices is implicitly assumed. In the following we will solve  Eqs.\,\eqref{eq:mainL2F} perturbatively.

\subsection{Perturbative solutions}\label{sec:pertcoord}

To solve the above~\eqref{eq:mainL2F} in the most general way, let us expand the component displacements perturbatively according to
\be
\label{eq:LPTweakexpansion}
  \fett{\xi}^\alpha = \fett{\xi}^{\alpha (1)} + \fett{\xi}^{\alpha (2)} + \ldots \,.
\ee 
In the following, after providing explicit perturbation equations for the component displacements,  we report the results for the fastest growing mode to first and second order. 

\paragraph*{First order.}
Formally linearizing the Lagrangian evolution equations, we obtain, to first order, a trivial identity from~\eqref{eq:LEOM2} implying that the fluid motion is potential in Lagrangian space. From Eq.\,\eqref{eq:LEOM1} we get at first order 
\begin{subequations}
\be \label{eq:1LPT2fluid}
  \RT \xi_{l,l}^{\alpha (1)} =   \frac{3g}{2D^2}  \left[ \fb\xi_{l,l}^{\rm b (1)} + \fc \xi_{l,l}^{\rc (1)}  \right]  ,
\ee
where we have used $q_i^{\rm c}(x_k^{\rm b}) = q_i^{\rm c}( q_l + \xi_l^{\rm b}) \simeq q_i - \xi_i^{\rm c} +  (\partial q_i^{\rm c}/\partial x_l) \xi_l^{\rm b} \simeq q_i - \xi_i^{\rm c}  + \xi_i^{\rm b}$ leading to $J^{\rm c}(q_l)|_{q_l = q_l^{\rm c}(x_k^{\rm b})} =  J^{\rm c} - J^{\rm c}_{,l} \xi_l^{\rm c} + J^{\rm c}_{,l} \xi_l^{\rm b} \simeq   1+ \xi_{l,l}^{\rm c(1)}$, and similarly for the term appearing in the evolution equation for component~c. 
To arrive at Eq.\,\eqref{eq:1LPT2fluid} we have used the boundary condition $\delta_\alpha (D_{\rm ini}) = \delta_\alpha^{\rm ini}$ (cf.~Eq.\,\ref{db}), which implies that  $\delta_\rM (D_{\rm ini}) \simeq 0$, a boundary condition that leads to Zel'dovich type solutions; see e.g.\ \cite{Buchert:1992ya,Rampf:2012xa}.
Of course, when slaving is applied which is the main focus in this paper, then $\delta_\rM(\text{\footnotesize $D=0$}) =0$ holds exactly.
 Furthermore, we have assumed that $\delta_\alpha^{\rm ini}$ are perturbations in the same sense as generic first-order displacement perturbations. For further calculational details, see Appendix~\ref{app:2fluid2LPT}.

The differential equation~\eqref{eq:1LPT2fluid} can be solved by considering difference  and weighted sum displacements which we define respectively as follows,
\be 
  \fett{\xi}^{\rM(1)} = \fb \fett{\xi}^{\rm b(1)} +\fc \fett{\xi}^{\rm c(1)}\,, \qquad  \fett{\xi}^{\rb \rc(1)} = \fett{\xi}^{\rm b(1)} - \fett{\xi}^{\rm c(1)}\,.
\ee 
At first order we then obtain from~\eqref{eq:1LPT2fluid} that 
\be
  \RT \xi_{l,l}^{\rm m(1)} =  \frac{3g}{2D^2} \xi_{l,l}^{\rM(1)} \,, \qquad \quad   \RT \xi_{l,l}^{\rm \rb\rc(1)} = 0 \,.
\ee
The general solutions of these first-order equations  can be analytically obtained for $g = 1$, leading to
\be
 \xi_{l,l}^{\rM(1)} = D\, C_+^{\xi} + D^{-3/2} C_-^{\xi}\,, \quad \xi_{l,l}^{\rb\rc (1)} = C_+^{\xi \rm bc} - 2 D^{-1/2} C_-^{\xi \rm bc} \,,
\ee
where the $C$'s are integration constants.
 Upon identification with the linearized Eulerian solutions, we have  $C_+^{\xi}= -C_+^{\rm m}$, $C_-^{\xi}= -C_-^{\rm m}$,  $C_-^{\xi \rb\rc} = - C_-^{\rb\rc}$, and $C_+^{\xi \rb\rc} =0$. From this one can  easily determine the general first-order solutions for the components $\alpha = \rb, \rc$, 
\begin{align} \label{eq:psibc1-twofluid}
 \begin{aligned}
   \xi_{l,l}^\rb &= - D \,C_+^\rM - D^{- 3/2}\, C_-^\rM   + 2 D^{-1/2} \fc \,C_-^\bc   \,, \\
   \xi_{l,l}^\rc &= - D \, C_+^\rM - D^{-3/2}\, C_-^\rM   - 2 D^{-1/2}  \fb \,C_-^\bc   \,. 
 \end{aligned}
\end{align}
\end{subequations}
We remark that setting $C_+^{\xi \rb\rc} =0$ stems from the present choice of Lagrangian coordinates, where  the corresponding mass conservation law receives an additional contribution $\sim \delta_\alpha^{\rm ini}$, which precisely compensates for the constant mode $C_+^{\xi \rb\rc}$. In \S\ref{sec:unpertcoord} we will introduce a set of perturbed Lagrangian coordinates that automatically incorporate terms $\sim \delta_\alpha^{\rm ini}$ in the displacement. 

The integration constants appearing in~\eqref{eq:psibc1-twofluid} still should be fixed. In the general case one should require that the component displacements vanish initially -- in accordance with the present definition of Lagrangian coordinates, together with appropriate conditions for their first time derivative (to specify the initial velocity). For growing-mode solutions, the first condition holds as well, while for the second condition one imposes the slaved boundary conditions~\eqref{eq:slaving2fluid} which effectively sets all decaying modes to zero and expresses all fields in terms of the initial gravitational potential. Doing so, we find for the growing mode of the two fluid components  
\be \label{eq:psi1lfuidsol-slaved}
  \boxed{ \fett{\xi}^{\alpha(1)}(\fett{q},D) = D\,\fett{\zeta}^{\rm m(1)}(\fett{q}) \,, }
\ee
with $\fett{\zeta}^{\rm m(1)}= - \nab \varphi^\ini$
which, due to the present choice of Lagrangian coordinates which assumes the mass conservation~\eqref{eq:massalphaLag}, and agrees exactly with the displacement in the single fluid case. 
We remark that growing-mode solutions can also be obtained in a more direct way; see \S\ref{sec:altgrowingL2F}.

\paragraph*{Second order.}
Next we consider the second-order perturbations. For this it is useful to define
the second-order invariants for arbitrary displacements~$\fett{\xi}$ and scalar $\varphi$
\begin{subequations}
\begin{align}
 & \mu_2^{\beta, \gamma} =  \frac 1 2 \left[ \xi_{i,i}^{\beta(1)} \xi_{j,j}^{\gamma(1)} - \xi_{i,j}^{\beta(1)} \xi_{j,i}^{\gamma(1)} \right] \,, \\
 & \mu_2(\varphi) = \frac 1 2 \left[(\varphi_{,ii})^2-(\varphi_{,ij})^2\right] \,, 
 \label{eq:mu2AB}
\end{align}
\end{subequations}
where, in the following,  $\beta$ and $\gamma$ are either m, b or c. 
Keeping only the second-order parts in Eq.\,\eqref{eq:LEOM1}, we find the following Lagrangian perturbation equations for the two-fluid system, 
\begin{subequations} \label{eq:EOM2lpt2lfuidsimple}
\begin{align}
   \RT   \xi_{l,l}^{\rm b(2)} \!& =\! \frac{3g}{2D^2} \Bigg[  \fb \! \left( \xi_{l,l}^{\rm b(2)} - \mu_2^{\rb,\rb} \right) +\fc \left( \xi_{l,l}^{\rm c(2)} +  \mu_2^{\rc,\rc} - 2 \mu_2^{\rm b,c}  \right) \nonumber \\ 
     & -\fc \bigg(  (\delta_{\rm c}^{\rm ini} - \xi_{i,i}^{\rm c(1)}) \partial_j  - \xi_{i,ij}^{\rm c(1)}  \bigg)  \bigg\{ \xi_{j}^{\rm b(1)} - \xi_{j}^{\rm c(1)} \bigg\}
\Bigg] \,, 
   \label{eq:EOM2lpt2lfuid1simplified} \\ 
\RT   \xi_{l,l}^{\rm c(2)} \!&=\! \frac{3g}{2D^2} \Bigg[ \fc \!\left( \xi_{l,l}^{\rm c(2)} - \mu_2^{\rc,\rc} \right)  + \fb \!\left( \xi_{l,l}^{\rm b(2)} + \mu_2^{\rb,\rb} - 2\mu_2^{\rm b,c} \right) \nonumber \\
   & + \fb  \bigg(   (\delta_\rb^{\rm ini} - \xi_{i,i}^{\rm b(1)}) \partial_j - \xi_{i,ij}^{\rm b(1)} \bigg)  \bigg\{ \xi_{j}^{\rm b(1)} - \xi_{j}^{\rm c(1)} \bigg\} \Bigg] \,, 
   \label{eq:EOM2lpt2lfuid2simplified}
\end{align}
while, using the first-order results~\eqref{eq:psibc1-twofluid} including decaying modes, the second-order part of~\eqref{eq:LEOM2} yields
\begin{align} \label{eq:cauchysol}
  \varepsilon_{ijk} \dot \xi_{j,k}^{\alpha(2)} =
  \varepsilon_{ijk} \bigg\{ & \frac 5 2 \hat C_{+,lj}^\rM \hat C_{-,lk}^\rM D^{-3/2} + 2 \hat C_{-,lj}^\rM \hat C_{-,lk}^{\rm bc} c^\alpha D^{-3}  \nonumber \\ 
 &- 3 \hat C_{+,lj}^\rM \hat C_{-,lk}^{\rm bc} c^\alpha D^{-1/2} \bigg\} \,,
\end{align}
\end{subequations}
where for the constants we employ a hat notation according to $\hat C = \nab^{-2} C$, and we have defined  $c^\alpha$ with $c^\rb = 1-\fb$ and $c^\rc = -\fb$. Calculational details to Eqs.\,\eqref{eq:EOM2lpt2lfuidsimple} are provided in Appendix~\ref{app:2fluid2LPT}.

\paragraph*{Fastest growing modes.} \ The second-order perturbation equations~\eqref{eq:EOM2lpt2lfuidsimple} can be easily integrated if needed. In the following, we report the results for the fastest growing modes, thus taking slaved boundary conditions into account. 
For this observe that in the absence of decaying modes, we have 
$\fett{\xi}^{\rm b(1)} = \fett{\xi}^{\rm c(1)}$ implying also
$\mu_2^{\alpha,\beta} = \mu_2^{\rM, \rM}$
and, of course $C_-^{\rm m} \to 0$,   $C_-^\bc \to 0$; thus, all  terms in curly brackets in the equations~\eqref{eq:EOM2lpt2lfuidsimple} vanish, which in particular implies for Eq.\,\eqref{eq:cauchysol} a vanishing source.
The evolution equations for the fastest growing modes thus simplify to
\begin{subequations} \label{eq:EOM2lpt2lfuidsimpleD}
\begin{align}
   &\RT   \xi_{l,l}^{\rm \alpha(2)} =
    \frac{3g}{2D^2} \Bigg[  \fb \xi_{l,l}^{\rm b(2)} + \fc \xi_{l,l}^{\rm c(2)} - D^2 \mu_2(\varphi^\ini) \Bigg]
    \,, \label{eq:EOM2lpt2lfuid1simplifiedD} \\
 & \varepsilon_{ijk} \dot \xi_{j,k}^{\alpha(2)} = 0 \,,\label{eq:EOM2lpt2lfuid1simplifiedDCauchy}
\end{align}
\end{subequations}
where $\mu_2$ is defined in~\eqref{eq:mu2AB}. Equation~\eqref{eq:EOM2lpt2lfuid1simplifiedDCauchy}
states the potential character for the first time derivative of the component displacements, which, by virtue of the used boundary conditions,  leads trivially to $\varepsilon_{ijk} \xi_{j,k}^{\alpha(2)} = 0$.
Equations~\eqref{eq:EOM2lpt2lfuidsimpleD} are easily solved, e.g., by employing the weighted sum and difference displacements
\begin{align}
  \fett{\xi}^{\rM(2)} = \fb \fett{\xi}^{\rb (2)} + \fb \fett{\xi}^{\rb (2)}\,, \qquad  \fett{\xi}^{\rb \rc(2)} = \fett{\xi}^{\rm b(2)} - \fett{\xi}^{\rm c(2)} \,,
\end{align}
which after suitable algebraic manipulations leads to\
\begin{subequations}
\begin{align}
  &\!\left[ \RT - \frac{3g}{2D^2} \right]   \xi_{l,l}^{\rM (2)} =  -\frac{3g}{2}  \mu_2(\varphi^\ini) \,, \\
 & \RT   \xi_{l,l}^{\bc (2)} = 0 \,, \label{eq:lastxibc2}
\end{align}
\end{subequations}
where $\RT$ is defined in Eq.\,\eqref{eq:Rop}. While Eq.\,\eqref{eq:lastxibc2} is merely a (perturbative)
repetition and thus is fixed to $\xi_{l,l}^{\bc (2)}=0$ with our choice of boundary conditions, the solution of the former
is the well-known second-order result in Lagrangian PT, i.e.,
\be
   \fett{\xi}^{\rM (2)}(\fett{q},D)  = - \frac{3D^2} 7 \, \nab^{-2}\,\nab  \mu_2(\varphi^\ini) \,,
\ee
which, of course agrees with the result given in the single-fluid \S\ref{sec:singlefluidLPT}. Since the second-order difference $\xi_{l,l}^{\bc (2)}$ is vanishing, the second-order growing-mode solution for the components are simply
\be \label{eq:xialpha2}
  \boxed{ \fett{\xi}^{\alpha(2)}(\fett{q},D) = \fett{\xi}^{\rm m(2)}(\fett{q},D) \,. }
\ee
We remind the reader that the results of this section are to be used with the mass conservation law~\eqref{eq:massalphaLag}, where the appearing $\delta_\alpha^\ini(\fett{q})$ is with our choice of coordinates and boundary conditions non-negligible.

Thus the component displacements formally agree  with  
the single fluid displacement (at least to third order, but see below), however only if the initial density perturbation $\delta_\alpha^{\rm ini}(\fett{q})$ is kept in the mass conservation law according to~\eqref{eq:massalphaLag}. Of course, that initial density perturbation affects
the density to all orders. Indeed, expanding~\eqref{eq:massalphaLag} to second order by using the results~\eqref{eq:psi1lfuidsol-slaved} and~\eqref{eq:xialpha2}, one finds firstly
\begin{align} \label{eq:massLPTalpha1}
  &\delta_\alpha(\fett{x}(\fett{q})) = \delta_\alpha^\ini(\fett{q}) + D \varphi_{,ll}^\ini(\fett{q}) + D \delta_\alpha^\ini \varphi_{,ll}^\ini \nonumber \\
  &\quad+ D^2 \left[ \frac 5 7 \varphi^\ini_{,ll} \varphi^\ini_{,mm} + \frac 2 7  \varphi^\ini_{,lm} \varphi^\ini_{,lm}  \right] 
     + O(3) \,.
\end{align}
For direct comparison with the Eulerian result, we need to evaluate all terms in the last expression at the identical (Eulerian) position,  for which we use the  ``pullback'' $\fett{q}(\fett{x}) = \fett{x}-\fett{\xi}$ to first order in
functions $F$ that depend on $\fett{q}$, i.e., $F(\fett{q}(\fett{x}))= F(\fett{x}-\fett{\xi}) = F(\fett{x}) - F_{,l} \xi_{,l}$ to first order.
As a consequence, the first two terms on the right-hand side of~\eqref{eq:massLPTalpha1} generate higher-order perturbations due to the pullback, leading to the ``Eulerian'' density
\begin{align} \label{eq:deltaLPT2fluidFirstApproach}
    &\delta_\alpha(\fett{x}) = \delta_\alpha^\ini(\fett{x}) + D \varphi_{,ll}^\ini(\fett{x}) + D \left( \delta_\alpha^\ini \varphi_{,ll}^\ini    + \delta_{\alpha,l}^\ini \varphi_{,l}^\ini \right) \nonumber \\
   &\quad+ D^2 \left[ \frac 5 7 \varphi^\ini_{,ll} \varphi^\ini_{,mm} + \varphi^\ini_{,llm} \varphi^\ini_{,m} +\frac 2 7  \varphi^\ini_{,lm} \varphi^\ini_{,lm}  \right] 
     + O(3) \,,
\end{align}
which agrees with the one obtained from the Eulerian calculation in \S\ref{sec:growing2SPT2F}. Furthermore, as shown in Appendix~\ref{app:2fluid2LPTvelocity}, the component velocity corresponding to the above reported component displacement evaluated at the Eulerian position agrees with the one from the Eulerian two-fluid result (Eq.\,\ref{eq:2SPT2fluids}).

Alternatively, for numerical applications such as generating initial conditions for simulations,  the initial densities can also be incorporated in the component displacements. As we will elucidate in \S\ref{sec:unpertcoord}, the resulting component displacements differ substantially from the above.

\subsection{Simplified derivation of growing-mode solutions}\label{sec:altgrowingL2F}

To obtain growing-mode solutions we can alternatively apply a similar simplification to the calculations as outlined in the Eulerian Section (see around Eq.\,\eqref{eq:ODEdeltaalphaPoissonsource}): for this we express the Poisson source of the shared two-fluid system by its counterpart in the single-fluid case. We thus replace in~\eqref{eq:LEOM1} the  term $(\nab_x^2 \varphi)_\alpha$ by the much simpler $\nab_x^2 \varphi(\fett{x}(\fett{q}))$, and express the latter by means of the single-fluid displacement~\eqref{eq:singleLPT} up to second order, i.e.,
\be \label{eq:nabla2Poisson}
   \nab_x^2 \varphi(\fett{x}(\fett{q})) = \frac{1/J(\fett{\xi}^\rM) -1}{D} =   \varphi_{,ll}^\ini + D \left[ \frac 5 7 ( \varphi_{,ll}^\ini)^2 +\frac 2 7 ( \varphi_{,lm}^\ini)^2 \right] \,. 
\ee
Doing so, Eq.\,\eqref{eq:LEOM1} becomes an ODE with 
an inhomogeneous term $\sim  \varphi_{,ll}^\ini$ and reads at first order 
\begin{align}
 \RT \xi_{l,l}^{\alpha(1)} = - \frac{3g}{2D} \varphi_{,ll}^\ini  \,. 
\end{align}
The analytic solution for non-decaying modes is $\xi_{l,l}^{\alpha(1)} = - D  \varphi_{,ll}^\ini + C_1^\alpha$ for $g= 1$. Here we have a choice as regards to the setting of $C_1^\alpha$. Assuming that mass
conservation for the components  is
\be \label{eq:masscomp}
  \delta_\alpha = \frac{1+ \delta_\alpha^\ini}{\det [\delta_{ij} + \xi_{i,j}^\alpha]} - 1 \,,
\ee
and expanding this to first order and set it equal to the Eulerian result for the component density (Eq.\,\ref{eq:deltaalpha2fl}), one easily establishes that $C_1^\alpha \to 0$ in this setting.  Similarly, if mass conservation is assumed to 
be  $\delta_\alpha = 1/\det [\delta_{ij} + \xi_{i,j}^\alpha] - 1$ then one finds that
$C_1^\alpha = - \delta_\alpha^\ini$; of course in that setting the Jacobian departs from unity already at initial time, contrary to what is assumed for~\eqref{eq:masscomp}. 
While both realizations are possible and consistent,
for the rest of the section we (continue to) assume mass conservation according to~\eqref{eq:masscomp} instead and thus set $C_1 = 0$,  while the other setting is effectively executed in the following subsection.

Iterating to the next order, we use again~\eqref{eq:nabla2Poisson} and obtain from Eq.\,\eqref{eq:LEOM1}
at second order
\be \label{eq:ODEinhomLPT}
  \RT \xi_{l,l}^{\alpha(2)} =  -\frac{15g}{7}  \mu_2(\varphi^\ini) \,,
\ee
where $\mu_2(\varphi^\ini)$ is defined in Eq.\,\eqref{eq:mu2AB}.
The non-decaying solution of Eq.\,\eqref{eq:ODEinhomLPT}
is $\xi_{l,l}^{\alpha(2)} = (-3/7) D^2 \mu_2$ where an occurring integration constant can be safely set to zero thanks to the boundary conditions and definition of the Lagrangian map.

Summing up, we find for the fluid components, truncated to second order in the growing modes, the simple result  
\be \label{eq:psi2lfuidsol-slaved}
  \fett{\xi}^\alpha(\fett{q},D) = D\,\fett{\zeta}^{\rm m(1)}(\fett{q}) + D^2 \fett{\zeta}^{\rm m(2)}(\fett{q}) \,,
\ee
where the spatial functions $\fett{\zeta}^{\rm m(1)}$ and $\fett{\zeta}^{\rm m(2)}$ are given in Eqs.\,\eqref{eq:sol2LPTsingle}.  

Actually, the above considerations carry over trivially to arbitrary high orders. Indeed, it is easily checked that when expressing the evolution equations for the component displacement in terms of the single-fluid displacement $\fett{\xi}$, i.e., 
\begin{align} \label{eq:mainL2Frep}
 \boxed{ 
   \begin{aligned}
      & \varepsilon_{ikl} \varepsilon_{jmn}  x_{k,m}^\alpha  x_{l,n}^\alpha \RT x_{i,j}^\alpha = - \frac{3g}{D^2} J^\alpha \left( \frac{1}{J(\fett{\xi}^\rM)} -1 \right) \,,  \\ 
      & \varepsilon_{ijk} x_{l,j}^\alpha \dot x_{l,k}^\alpha =0 \,, \\ 
   \end{aligned}
 }
\end{align}
and uses the known recursion relations for $\fett{\xi}^\rM = \sum_{n=1}^\infty \fett{\zeta}^{\rM (n)} \,D^n$ \citep[see e.g.][]{Zheligovsky:2014}, then one establishes the following simple all-order result for all coefficients $1\leq n  < \infty$ of the component displacement,
\be \label{eq:allorderdispl2f}
 \boxed{  \fett{\xi}^{\alpha (n)} =  \fett{\xi}^{\rM (n)} \,.}
\ee
We have explicitly verified that the simplified equations~\eqref{eq:mainL2Frep} provide identical results with the approaches of \S\ref{sec:growing2SPT2F} and~\ref{sec:pertcoord} up to third order; see Appendix~\ref{app:sumdifference-nonpert} for details. Furthermore, in Appendix~\ref{app:xibc} we recursively prove that
\be
 \fett{\xi}^{\bc (n)} \equiv \fett{\xi}^{\rb (n)} -  \fett{\xi}^{\rc (n)}=0 \,,
\ee
from which it follows that Eq.\,\eqref{eq:allorderdispl2f} actually holds at all orders.

\subsection{Incorporating initial density perturbations in component displacements}\label{sec:unpertcoord}

We have seen that for the two-fluid case with growing-mode solutions, the initial density perturbations
$\delta_\alpha^{\rm ini}$ are non-negligible; ignoring them would induce quasi-singular behaviour for $D \to 0$. This should be contrasted to similar derivations in the single-fluid case, where initial density perturbations are usually ignored; for a discussion see e.g.~\cite{Rampf:2012xa}. We note however that initial density perturbations can also not be ignored in Lagrangian bias expansions since biased/observed objects are generally not uniformly distributed, see e.g.\ \cite{Matsubara:2011ck,Desjacques:2016bnm}.

Recall that $\delta_\alpha^{\rm ini}$ appears in the mass conservation for the components, Eq.\,\eqref{eq:masscomp},
which is to be used with the corresponding map~$\fett{x}^\alpha$ given in Eq.\,\eqref{eq:psi2lfuidsol-slaved}.
The presence of nonzero $\delta_\alpha^{\rm ini}$ acts as a space-dependent distortion,  which can be treated in several ways. For example, numerical simulations could be initialized with a varying $N$-body particle mass to reflect the local density distortion.

An alternative -- which we employ here -- is to introduce distorted Lagrangian coordinates that automatically absorb $\delta_\alpha^{\rm ini}$.  To absorb $\delta_\alpha^{\rm ini}$ into the displacements, we consider the two consecutive ``maps''
\be
  \fett{y}_\alpha(\fett{q}) = \fett{q} + \nab \chi_\alpha(\fett{q}) \,,\qquad 
  \fett{x}_\alpha(\fett{q}, D) = \fett{q} + \fett{\xi}_\alpha(\fett{q},D) \,,
\ee
where $\nab \chi_\alpha$ is a {\it time-independent yet non-perturbative displacement} generated by the initial density perturbation $\delta_\alpha^\ini$, and the component displacements~$\fett{\xi}_\alpha$ are given in Eq.\,\eqref{eq:psi2lfuidsol-slaved}.
To evolve the fluid system for these two consecutive maps, we use the composition
\begin{align}
  \fett{x}_\alpha^{\rm full}(\fett{q}) &\equiv  \fett{x}_\alpha \circ \fett{y}_\alpha 
     = \fett{x}_\alpha(\fett{y}_\alpha(\fett{q})) = \fett{x}_\alpha(\fett{q}+ \nab\chi_\alpha) \nonumber \\
   & =  \fett{q} + \fett{\xi}_{\alpha}  +\fett{\xi}_{\alpha,l} \chi_{\alpha,l} + \nab \chi_\alpha + O(3)  \,, \label{eq:xfullcomp}
\end{align}
such that mass conservation reads exactly 
\be
  \delta_\alpha (\fett{x}^{\rm full}(\fett{q},D)) \equiv \frac{1}{\det [x_{\alpha i,j}^{\rm full}(\fett{q},D) ]} - 1\,,
\ee 
i.e., $\delta_\alpha^{\rm ini}$ is absorbed in $\fett{x}^{\rm full}$. At the same time, the initial density expressed in terms of the $\fett{y}_\alpha$-map is governed by 
\be \label{eq:deltainiY}
  \delta_\alpha^\ini (\fett{y}^{\rm full}(\fett{q}))  \equiv \frac{1}{\det [ y_{\alpha i,j}(\fett{q})]} - 1\,, 
\ee
where $\det [ y_{\alpha i,j}] = 1 + \nab^2 \chi_\alpha + \mu_2(\chi_\alpha)  + \det[\chi_{\alpha i,j}]$, where $\mu_2(\chi_\alpha)$ is defined in Eq.\,\eqref{eq:mu2AB}. To proceed we need to determine the displacement $\nab \chi_\alpha$ for the  $\fett{y}_\alpha$-map to the same accuracy level as employed for the $\fett{x}_\alpha$-map. We thus assume the perturbative expansion $\chi_\alpha = \chi_\alpha^{(1)} + \chi_\alpha^{(2)}$, for which~\eqref{eq:deltainiY} becomes 
\be
  \delta_\alpha^\ini (\fett{y}^{\rm full}(\fett{q}))  =  - \chi_{\alpha,ll}^{(1)}(\fett{q}) -  \chi_{\alpha,ll}^{(2)} - \mu_2(\chi_\alpha^{(1)}) + (\nab^2 \chi_\alpha^{(1)})^2 
\ee
to second-order accuracy.
Clearly, the first-order solution is simply 
\be
 \chi_\alpha^{(1)}(\fett{q}) = -\nab^{-2} \delta_\alpha^\ini(\fett{y}^{\rm full}(\fett{q})) \simeq -\nab^{-2} \delta_\alpha^\ini(\fett{q}) \equiv \zeta_\alpha^\ini(\fett{q}) \,,
\ee 
where we have used the fact that the initial density perturbation is, to the leading order, evaluated at the position $\fett{q}$. At second order, however, the actual functional dependence of $\delta_\alpha^\ini (\fett{y}^{\rm full}(\fett{q})=\delta_\alpha^\ini (\fett{q}+ \nab \chi_\alpha)$ must be taken into account, which generates a second-order term -- similarly as discussed in the previous section.
It is then straightforward to determine $\chi_\alpha$; truncated to second order it reads 
\be
 \chi_\alpha = - \nab^{-2} \left[ \delta_\alpha^\ini(\fett{q}) + \mu_2(\zeta_\alpha^\ini) + (  \delta_\alpha^\ini \zeta_{\alpha,l}^\ini )_{,l}  \right] + O(3) \,.
\ee
Using this in~\eqref{eq:xfullcomp} we finally obtain the composite map 
\begin{align} \label{eq:comm-prosite-map}
 \boxed{
   \begin{aligned}   \fett{x}_\alpha^{\rm full}  &= \fett{q} + D\,\fett{\zeta}^{\rm m(1)}(\fett{q}) + D^2 \fett{\zeta}^{\rm m(2)}(\fett{q}) 
     - D \zeta_{\alpha,l}^{\rm ini} \fett{\nabla} \varphi_{,l}^{\rm ini}  \\ 
  & \quad +  \fett{\nabla} \zeta_\alpha^{\rm ini} -  \nab^{-2} \fett{\nabla} ( \delta_\alpha^\ini \zeta_{\alpha,l}^{\rm ini})_{,l}   - \nab^{-2} \fett{\nabla} \mu_2(\zeta_\alpha^\ini) \,, 
   \end{aligned}
 }
\end{align}
with $\zeta_\alpha^\ini = -\nab^{-2} \delta_\alpha^\ini(\fett{q})$, whereas the spatial functions~$\fett{\zeta}^{\rm m(1)}(\fett{q})$ and~$\fett{\zeta}^{\rm m(2)}(\fett{q})$ are given in Eqs.\,\eqref{eq:sol2LPTsingle}.
Using the methods described above, it is easily shown that this composite map exactly produces the same second-order component density (Eq.\,\ref{eq:deltaLPT2fluidFirstApproach}) as in the approaches of Sections~\ref{sec:growing2SPT2F} and~\ref{sec:altgrowingL2F}. 
Furthermore, in Appendix~\ref{app:2fluid2LPTvelocity} we show that the associated velocity at the current (Eulerian) position agrees as well with the Eulerian result.

In summary, on a theoretical footing, the method presented here agrees with the one of \S\ref{sec:altgrowingL2F}. For the numerical application, the methods are however fairly distinct: While for the method of \S\ref{sec:altgrowingL2F} we should incorporate the initial density perturbation by varying the particles masses in an $N$-body simulation, no such thing is necessary in the present method. On the downside, incorporating the initial density perturbations in the component displacements can only be done order by order, while the method of~\S\ref{sec:altgrowingL2F} applies directly to all orders.  Furthermore, incorporating the initial density in the displacements leads to large discreteness errors in the numerical solutions; for details see  Section~4 in our companion~paper.

\section{Variational approach for LSS in \texorpdfstring{$\fett{\Lambda}$}{L}CDM} \label{sec:H&HJE}

Complementary to the Eulerian and Lagrangian fluid approaches, recently a semi-classical formalism has been put forward by \cite{Uhlemann:2019}. This formalism is related to a classical Hamiltonian theory, however only once the corresponding momentum variable is non-canonically transformed -- this is the essence that leads to a direct relationship between Hamiltonian theory and the cosmological fluid description \citep[see e.g.][]{Bartelmann:2014fma}. Here we reconsider this relationship and show that the classical Hamiltonian can be transformed to a ``new'' (contact) Hamiltonian that lives on a special manifold on an extended phase-space. 
We will show that, on the one hand, perturbative solution techniques are amenable to this transformed Hamiltonian -- which is not at all straightforward within the classical Hamiltonian theory. On the other hand, this
 Hamiltonian builds the basis for the semi-classical approach discussed in \S\ref{sec:schroedi}, whose perturbative solutions in terms of the propagator are related to the classical action along the phase-space trajectory.

We begin with the Hamiltonian in $\Lambda$CDM  which is 
\be
\label{eq:Hcosmic}
 \tilde H (\fett{x},\fett{p}, t) = \frac{{\fett{p}}^2}{2ma^2(t)} + \tilde \varphi(\fett{x})\,,
\ee 
where $\fett{p} = m a^2 \tilde {\fett{u}}$ is the canonically conjugated momentum, and
 we have employed the same (tilde) notation as in \S\ref{sec:singleEuler}.
To change the time variable from cosmic time to $D$-time, recall that 
$\dd/\dd t = (\partial_t D) \dd / \dd D$, and consider the action
\be
  {\cal S}_{\rm p} = \!\int \!\!{\cal L} \dd t   = \int \left( \fett{p} \cdot m \frac{\dd \fett{x}}{\dd D} - \frac{\fett{p}^2}{2 m a^2 (\partial_t D)} - \frac{3D}{2a (\partial_t D)}  \varphi  \right) \dd D\,,
\ee
where $\tilde \varphi = 3D \varphi/(2a)$, with $\varphi$ governed by the Poisson equation~\eqref{eq:PoisD}. From the action  we can read off the Hamiltonian in  $D$-time, 
\be \label{eq:HinDtime}
   H(\fett{x},\fett{p}, D) =  \frac{\fett{p}^2}{2 m a^2 (\partial_t D)} + \frac{3D}{2a (\partial_t D)} \varphi(\fett{x}) \,.
\ee
It is easily seen that the corresponding Hamiltonian evolution equations are incompatible with slaving (cf.\ \S\ref{sec:slaving}),  which, on a technical level stems from the fact that in a Hamiltonian theory one is forced to employ canonically conjugate variables. Liouville's theorem essentially forces momentum space to expand indefinitely, while comoving coordinate space contracts to a point as $a\to0$.  To proceed, and to make the connection to the fluid approach more transparent, one may express the Hamiltonian in terms of the velocity $\fett{v}$ instead of $\fett{p}$ which is, strictly speaking, a non-canonical transformation. 

An ``alternative'' of this is making use of the  so-called contact geometry, where the above Hamiltonian can be ``contact transformed'' -- which is a generalization of canonical transformations; see e.g.\ \cite{2001dssg.book.....A,BRAVETTI201717}. 
Historically, the early development of contact geometry traces back to Sophus Lie \citep[cf.][]{Etnyre}, and was later on applied to geometrize thermodynamics. 
More recently, contact geometry has been employed in Hamiltonian dynamics, with important fundamental work performed by Vladimir Arnold in the late 80s.
In the following we shall employ the contact approach, while calculational details and further results are provided in Appendix~\ref{app:HJ}.

Expressing~\eqref{eq:HinDtime} in terms of the velocity $\fett{v}$ (we set $m=1$) {\it which is here an independent variable and the contact conjugate to} $\fett{x}$, 
we obtain the contact Hamiltonian in $\Lambda$CDM, derived in Appendix~\ref{app:HJ}, 
\begin{subequations} 
\be \label{eg:contactH}
 \boxed{ \! {\cal H}(\fett{x}, \fett{v}, {\cal S}, D) = \frac{\fett{v}^2}{2} +  {\cal V} , \,\,\,\,\, {\cal V} (\fett{x},{\cal S},D)= \!\frac{3g}{2D} \left( \varphi (\fett{x}) + {\cal S} \right) , \!}
\ee 
where ${\cal S} = \int \fett{v} \cdot \dd \fett{x} - {\cal H} \dd D$ is the corresponding action (up to an integration constant). 
We remark that the corresponding Hamiltonian equations of motion for~\eqref{eg:contactH} are actually independent of ${\cal S}$, they read (see Appendix~\ref{app:HJ})
\begin{align} \label{eq:contactEOMs}
   \frac{\dd \fett{x}}{\dd D}  = \fett{v}  \,, \qquad \quad   \frac{\dd \fett{v}}{\dd D}  &= - \frac{3g}{2D} \left( \fett{v} + \nab \varphi \right) 
\end{align}
\end{subequations}
for the position and velocity variables, while ${\cal S}$ is determined through
$\dd {\cal S}/\dd D = \fett{v}^2/2 - 3g [ \varphi  +  {\cal S}]/(2D)$.
In \S\ref{sec:schroedi} we show how the contact Hamiltonian relates to a Schr\"odinger equation.

Now, defining the generating function ${\cal W}(\fett{x}', D'; \fett{x}, D)$, which physically corresponds to the action along the phase-space trajectory from $(\fett{x}', D')$  to  $(\fett{x}, D)$, with 
\be \label{eq:generatingfunc}
  {\cal S}_\fixedxptp : (\fett{x}, D) \mapsto {\cal W}(\fett{x}', D'; \fett{x}, D)  = \int_{\fett{x}', D'}^{\fett{x}, D} \fett{v} \cdot \dd \fett{x} - {\cal H} \dd D \,,
\ee
we find that the generating function is governed by the Hamilton--Jacobi equation
\be \label{eq:HJ}
   \boxed{ \partial_D {\cal S}_\fixedxptp +  {\cal H}(\fett{x}, \nab_x {\cal S}_\fixedxptp \,, {\cal S}_\fixedxptp \,, D) = 0 \,, }  
\ee 
for fixed (initial) coordinates $(\fett{x}', D')$.
Using $\nab_x {\cal S}_\fixedxptp$ as the velocity variable confines the phase-space trajectories on the so-called Legendrian submanifold  \citep[cf.][]{2000JMP....41.3344E}, which is the contact analogue to the Lagrangian submanifold in standard symplectic geometry, and thus of importance when determining the phase-space of infinitely cold matter \citep[see e.g.][]{Abel:2012}.

The Hamiltonian~\eqref{eg:contactH}, as well as the Hamilton--Jacobi equation~\eqref{eq:HJ} comprise individual starting points  for investigating perturbative solutions in phase-space. For this it is important to note that both~\eqref{eg:contactH} and~\eqref{eq:HJ} remain regular at $D=0$ provided we use the slaving conditions
\be
\label{eq:slaveS}
 \lim_{D' \to 0} {\cal S}_\fixedxptp 
    \equiv {\cal S}^{\rm ini} (\fett{x}') =  -\varphi^{\rm ini} \,,
\ee
which, as in the previous sections, should be supplemented with the statement of initial quasi-homogeneity. 
It is also worthwhile to point out that the Hamilton--Jacobi equation~\eqref{eq:HJ} 
admits solutions to the corresponding Cauchy problem \cite[cf.][]{doi:10.1142/10307}, it reads
\be \label{eq:CauchyProb}
  {\cal S}_{\text{\scriptsize $\fett{x}', D'=0$}\,} (\fett{x},D)= {\cal S}^{\rm ini}(\fett{x}') + {\cal W}(\fett{x}',\fett{x}; 0, D)\,,
\ee
which, in the cosmological case, should remain meaningful at least until the instance of the first shell-crossing.
Perturbative solutions in terms of the growing-mode $D$ are discussed in the following, while the inclusion of decaying modes will be kept for future work.

\subsection{Single-fluid case}\label{sec:singleHJE}

As motivated above, the generating function ${\cal W}$ is governed by the Hamilton--Jacobi equation
\be \label{eq:HJrep}
  \partial_D {\cal W}(\fett{q}, 0; \fett{x}, D) +  \frac{[\nab_x {\cal W}(\fett{q}, 0; \fett{x}, D])^2}{2} + {\cal V} = 0 \,,
\ee
where, instead of $\fett{x}'$, from here on we use $\fett{q}$ to denote the initial (fixed) coordinate.
From conventional PT, which is valid before shell-crossing, we know that, to the leading order and with the appropriate choice of coordinates, the motion of fluid particles in an expanding Universe is ballistic with prescribed initial velocities.
At the level of the Hamiltonian~\eqref{eg:contactH}, this statement implies that ${\cal V} \simeq 0$ to first order, which translates into the following potential-free Hamilton--Jacobi equation, 
\be
   \partial_D {\cal W}_{\rm free} + (\nab_x {\cal W}_{\rm free})^2/2 = 0  \,.
\ee
It is elementary to solve such an equation within the context of classical and quantum mechanics \citep[see e.g.][]{doi:10.1142/10307}, leading to 
\be \label{eq:freeW}
   {\cal W}_{\rm free}(\fett{q},0;\fett{x}, D) = \frac{(\fett{x}-\fett{q})^2}{2D} \,,
\ee
the so-called free-particle generating function.

To incorporate ${\cal V} \neq 0$ and thus to solve the Hamilton--Jacobi equation~\eqref{eq:HJrep}
to second-order accuracy,
we adopt the methodology of~\cite{doi:10.1142/10307} and impose a power-law {\it Ansatz} for the generating functional
\be
   {\cal W}(\fett{q},0;\fett{x}, D) = {\cal W}_{\rm free}(\fett{q},0;\fett{x}, D) + \sum_{n=2}^\infty {\cal W}^{(n-1)} (\fett{q},\fett{x})\, D^{n-1} \,,
\ee
which is an eligible {\it Ansatz} 
provided that ${\cal V}$ can be represented in terms of a power series in $D$ -- as it is the case in PT where ${\cal V} = \sum_{n=1} {\cal V}^{(n)} D^{n-2}$ \citep{Uhlemann:2019}.
Plugging this into~\eqref{eq:HJrep} together with the leading-order result~\eqref{eq:freeW}, 
we obtain the following condition, at $n=2$,
\be \label{eq:Wconstraint}
 {\cal W}^{(2)} + (\fett{x} -\fett{q}) \cdot \nab_x  {\cal W}^{(2)} + {\cal V}^{(2)} =0 \,,
\ee
which is solved exactly by the following second-order result
\be
\label{eq:Wsol}
  {\cal W}^{(2)}(\fett{q},\fett{x}) = - \int_0^1  {\cal V}^{(2)} \left(\fett{q} + s [\fett{x} - \fett{q}] \right) \dd s \,.
\ee
For practical applications, such as for the Schr\"odinger approach employed in \S\ref{sec:schroedi}, one may also use the two-endpoint approximation for the above, which is $ {\cal W}^{(2)} \simeq - \left[ {\cal V}^{(2)}(\fett{q}) + {\cal V}^{(2)}(\fett{x})  \right]/2$.

Observe that the above result for the generating functional, which is essentially a result of the Cauchy problem~\eqref{eq:CauchyProb}, does not require explicit solutions of the Poisson equation. In fact this independence is one of the known key advantages of Hamilton--Jacobi. By contrast, in the Hamiltonian approach any progress -- be it analytical or numerical, requires either solving for the equations motion, or by considering invariants which are constants of motion along the path; thus, either way, the Hamiltonian approach requires the explicit knowledge of  the Poisson equation.

Alternatively to a full-fledged PT in phase-space, one may also employ conventional PT to express some of the required functions approximatively. For example, the Hamiltonian~\eqref{eg:contactH} valid to second-order within a $D$ expansion is
\be \label{eq:Hsingle}
   {\cal H} = \frac{\fett{v}^2}{2} +  V_{\rm eff}^{(2)} \,, \qquad V_{\rm eff}^{(2)} = \frac 3 7 \nab^{-2}\! \left[  \varphi_{,ll}^{\rm ini} \varphi_{,mm}^{\rm ini} -  \varphi_{,lm}^{\rm ini} \varphi_{,lm}^{\rm ini}  \right] \,.
\ee
Calculational details about how $V_{\rm eff}$ is determined are given in Appendix~\ref{app:2fluid2SPT}.

\subsection{Two-fluid case}\label{sec:2fluidHJ}

Consider now the two-fluid (contact) Hamiltonian
\be \label{eq:contactHalpha}
   {\cal H}(\fett{x}_\alpha , \fett{v}_\alpha , {\cal S_\alpha }, D) = \frac{\fett{v}_\alpha^2}{2} + \frac{3g}{2D} \left[  {\cal S}_\alpha + \varphi(\fett{x}_\alpha)  \right]\,,
\ee
where the common potential is governed by the Poisson equation 
\begin{align} \label{eq:poissonmulti}
 &\left( \nab^2_x \varphi \right)_\alpha  =  \Bigg[ \int \Big\{  \fb [ 1+\delta_\rb^\ini(\fett{q})] \,\delta_{\rm D}^{(3)} (\fett{x}^\alpha - \fett{x}_\rb (\fett{q}, D))  \nonumber \\
    &\qquad + \fc  [1+\delta_\rc^\ini(\fett{q})]\, \delta_{\rm D}^{(3)} (\fett{x}^\alpha - \fett{x}_\rc (\fett{q}, D)) \Big\} \dd^3 q - 1\Bigg] \Big/ D
\end{align}
(see~e.g.\ \citealt{Pietroni:2018,Rampf:2019nvl} for applications of this expression to the single fluid case, and e.g.~\citealt{Chen:2019cfu} for the two-fluid case),
where $\delta_{\rm D}^{(3)}$ is the Dirac delta and we have introduced  the sought solutions in phase-space $\fett{x}_\rc (\fett{q}, D)$ in parametric form.
To guarantee regularity at arbitrary short times we should  impose
that Eq.\,\eqref{eq:poissonmulti} remains finite for $D\to 0$, as well as demand the following slaving condition on the individual generating functionals
\be
  \lim_{D' \to 0} {\cal S}_{\fett{x}', D'}^\alpha  
    \equiv {\cal S}_\alpha^{\rm ini} (\fett{x}') =  -\varphi^{\rm ini} \,,
\ee
where ${\cal S}_{\fett{x}', D'}^\alpha$ is equivalently defined as in the single fluid case (cf.\ Eq.\,\ref{eq:slaveS}).

Similarly as in the single-fluid case these equations could be solved perturbatively by using standard methods known from symplectic geometry. However, for the purpose for applying the above to a phenomenological Schr\"odinger equation (\S\ref{sec:2fluidschroedi}), it is useful to determine the two-fluid  Hamiltonian~\eqref{eq:contactHalpha} valid to second-order accuracy in the strict $D$-expansion. Calculations details for determining the effective potential $V_{\rm eff}^{(2)}$ are provided in  Appendix~\ref{app:2fluid2SPT}, leading to the Hamiltonian for fluid component $\alpha=\rb,\rc$, 
\begin{subequations}
\be
 \boxed{  {\cal H}(\fett{x}_\alpha,\fett{v}_\alpha,D) = \frac{\fett{v}_\alpha^2}{2} +  V_{\rm eff}^{(2)}(\fett{x}_\alpha)\,,  }
\ee
where 
\be \label{eq:Veff}
   V_{\rm eff}^{(2)}(\fett{x}) = \frac 3 7 \nab^{-2}\! \left[  \varphi_{,ll}^{\rm ini} \varphi_{,mm}^{\rm ini} -  \varphi_{,lm}^{\rm ini} \varphi_{,lm}^{\rm ini}  \right] \,,
\ee
\end{subequations}
to second-order accuracy.

\section{Semi-Classical approach for LSS in \texorpdfstring{{$\fett{\Lambda}$}}{L}CDM} \label{sec:schroedi}

The perturbative contact Hamiltonian~\eqref{eq:HinDtime} discussed in the previous section can be used to formulate a Schr\"odinger wave equation.
Based on this, we generalize the semi-classical approach from \cite{Uhlemann:2019} \citep[an extension of the free-particle approximation from][]{Short:2006md,Short:2006me}  for a single fluid in EdS to two fluids in $\Lambda$CDM. In so-called propagator perturbation theory (PPT), we obtain perturbative solutions for the wave function, which straightforwardly predict Eulerian fluid observables while simultaneously implementing a semi-classical analogue of Lagrangian PT.

Before proceeding, we remark that formulating a wave equation using  
the canonical Hamiltonian~\eqref{eq:Hcosmic} would lead to the Schr\"odinger-Poisson equation describing structure formation for fuzzy dark matter (see \citealt{Hui:2017} for a review), or an approximate treatment of standard cold dark matter \citep[see e.g.][]{Widrow:1993,Uhlemann:2014,2017PhRvD..96l3532K}. Here we do not follow this idea because formulating a PT for the Hamiltonian~\eqref{eq:Hcosmic} is hampered by the distinct time-dependencies of the kinetic and potential terms in the Hamiltonian. Furthermore, expanding the wave function in amplitude and phase, as done in wave PT presented by \cite{Li:2019}, appears to be even more limited than Eulerian PT.

\subsection{Single-fluid case}

The wave function analogue of the fluid equations for a single fluid is obtained as a solution to the Schr\"odinger equation
\begin{subequations} \label{eq:S1fluid}
\begin{align} 
\label{eq:Schroedi1fluid}
 \ii \hbar \partial_D \psi(\fett{x},D) &= \hat {\cal H} \psi(\fett{x},D) \,, \quad\, \hat {\cal H} =  -\frac{\hbar^2 \nab^2_x}{2}  + \! \frac{3g}{2D} \left( {\cal S} + \varphi \right) , 
\end{align}
supplemented with the Poisson equation
\be
  \nab^2_x \varphi = \frac{|\psi|^2-1}{D} \,.
\ee
\end{subequations}
Here, $\hat {\cal H}$ may be viewed as the contact-Hamiltonian given in Eq.\,\eqref{eq:Hsingle} in operational form, where ${\cal S}$ is the solution of a Bernoulli-type equation (see Eq.\,\eqref{eq:actionD}), and is thus intrinsically associated with the velocity potential (before shell-crossing). 

Before investigating the non-linear Schr\"odinger theory, it is useful to establish the necessary boundary conditions for growing-mode solutions. It is easily verified that Eqs.\,\eqref{eq:S1fluid} remain regular for $D \to 0$ provided we use the slaving conditions, which in the present context translate to
\be
  \left| \psi^\ini \right|^2 = 1\,, \qquad {\cal S}^\ini = - \varphi^\ini \,,
\ee
where ``ini'' denotes as before evaluation at $D=0$. These slaving conditions imply for the wave function initially
\be
 \label{eq:psi_ini1}
\psi^{\rm ini}(\fett{q}) = \exp \left[ \frac{\ii}{\hbar} \varphi^{\rm ini}(\fett{q})\right]\,,
\ee
where, in accordance with our used notation in this paper, $\fett{q}$ can be viewed as the initial particle coordinate.
Supplemented with this initial condition,  \cite{Uhlemann:2019} suggested  to solve the Schr\"odinger equation by employing the {\it propagator perturbation theory} (PPT) which has at its central object the 
propagator $K(\fett{q},0 \,|\, \fett{x},D) =: K(\fett{q}, \fett{x}; D)$, which propagates a wave function $\psi$ from its initial state to the final (or current) state at time $D$, i.e., 
\be
\label{eq:PsiKernel}
  \psi(\fett{x};D) =  \int \! \dd^3 q \, K(\fett{q},\fett{x};D) \,  \psi^{\rm ini}(\fett{q}) \,.
\ee
Knowing the propagator and the initial conditions implies knowing the solution to the wave function.

In PPT, the quantity $V_{\rm eff} \equiv 3g ({\cal S} + \varphi)/(2D)$ is taken to be an external potential, which can be easily determined using conventional PT. 
The Schr\"odinger equation~\eqref{eq:S1fluid} then becomes
\be \label{eq:S2}
  \ii \hbar \partial_D \psi(\fett{x},D) =  \left[ -\frac{\hbar^2 \nab^2_x}{2}  +  V_{\rm eff}(\fett{x}, D) \right] \psi(\fett{x},D) \,, 
\ee
with $V_{\rm eff}^{(1)}\equiv 0$ and $V_{\rm eff}^{(2)}(\fett{x}) =  (3/7) \nab^{-2}\left[  \varphi_{,ll}^{\rm ini} \varphi_{,mm}^{\rm ini} -  \varphi_{,lm}^{\rm ini} \varphi_{,lm}^{\rm ini}  \right]$, which are evidently time-independent at the considered orders.
In the following we solve Eq.\,\eqref{eq:S2} using PPT.

\paragraph*{Leading order.}
Since $V_{\rm eff}^{(1)}=0$, the Schr\"odinger equation~\eqref{eq:Schroedi1fluid} is potential free at the leading order, i.e., 
\be
  \ii \hbar \partial_D \psi_{\rm free}(\fett{x},D) = - \frac{\hbar^2}{2} \nab_x^2 \psi_{\rm free}(\fett{x},D) \,.
\ee
Solving for the associated propagator $K_{\rm free}$ for the free wave function~$\psi_{\rm free}$, one easily finds
\be
\label{eq:K0sol}
 K_{\text{\scriptsize free}}(\fett{q},\fett{x};D) =  (2 \pi \ii \hbar D)^{-3/2}
   \exp\left[ \frac{\ii}{\hbar} \frac{(\fett{x}-\fett{q})^2}{2D} \right] \,,
\ee
where the normalization factor is introduced such that the propagator amounts to the Dirac delta~$\delta_{\rm D}^{(3)}(\fett{x}-\fett{q})$  for $D \to 0$, and thus, Eq.\,\eqref{eq:PsiKernel} returns the initial wave function~\eqref{eq:psi_ini1} at $D=0$.

Observe that, apart from the prefactor $\ii/\hbar$, the exponential in~\eqref{eq:K0sol} is nothing but the free generating functional ${\cal W}_{\rm free}$, Eq.\,\eqref{eq:freeW},  in the Hamilton--Jacobi approach.

\paragraph*{Beyond leading order.} Beyond the leading order, the effective potential $V_{\rm eff}^{(2)}$ needs to be included in the analysis.
 To iterate to arbitrary high orders, we impose the {\it Ansatz} for the propagator 
\be \label{eq:Ansatzprop}
  K(\fett{q},\fett{x};D) = \! K_{\text{\scriptsize free}}(\fett{q},\fett{x};D)  \exp\left[ \frac \ii \hbar \sum_{n=2}^\infty \! {\cal M}^{(n-1)}(\fett{q},\fett{x})\, D^{n-1}  \right].
\ee
The first new unknown ${\cal M}^{(2)}$ is easily determined by first plugging the {\it Ansatz}~\eqref{eq:Ansatzprop} into the Schr\"odinger equation~\eqref{eq:Schroedi1fluid} and keeping only terms at second order in PPT, which leads to the equation
\begin{subequations}
\be
  {\cal M}^{(2)} + (\fett{x} -\fett{q}) \cdot \nab_x  {\cal M}^{(2)} + V_{\rm eff}^{(2)} =0 \,,
\ee
with solution
\be \label{eq:solK1}
 {\cal M}^{(2)}(\fett{q},\fett{x}) = - \int_0^1  V_{\rm eff}^{(2)} \left(\fett{q} + s [\fett{x} - \fett{q}] \right) \dd s \,,
\ee
\end{subequations}
which coincides precisely with the corresponding second-order solution for the generating functional ${\cal W}^{(2)}$ in equation~\eqref{eq:Wsol} in the Hamilton--Jacobi approach (\S\ref{sec:singleHJE}).
\cite{Uhlemann:2019} advocated to use the two-endpoint approximation for the 2PPT kernel~\eqref{eq:solK1},  
for which the propagator $K =  K_{\text{\scriptsize 2PPT}}$ becomes
\be
\label{eq:KNLOsol}
 \boxed{ K_{\text{\scriptsize 2PPT}} (\fett{q},\fett{x};D) = K_{\text{\scriptsize free}}(\fett{q},\fett{x};D)  \exp\left[ -\frac{\ii D}{2\hbar}\left[ V_{\rm eff}^{(2)}(\fett{q}) + V_{\rm eff}^{(2)}(\fett{x}) \right] \right] ,}
\ee
implying that the effective potential~\eqref{eq:Veff} is evaluated at the initial and final positions which resembles a numerical kick-drift-kick scheme; see~\cite{Uhlemann:2019} for details 
and \cite{HahnAccomp2020} for an application of this method to initial conditions for $N$-body simulations.

\subsection{Two-fluid case}\label{sec:2fluidschroedi}

For two gravitationally coupled fluids, the component Schr\"odinger equation reads ($\alpha=\rb,\rc$)
\begin{subequations}
\label{eq:Schroedi2}
\begin{align} 
\label{eq:Schroedi2fluid}
 \ii \hbar \partial_D \psi_\alpha &= \ - \frac{\hbar^2}{2} \nab_x^2 \psi_\alpha + \frac{3 g}{2D} \big( {\cal S}_\alpha +  \varphi \big) \psi_\alpha\,,
\end{align}
which is to be supplemented with the Poisson equation
\be
  \nab_x^2 \varphi = \frac{\fb |\psi_\rb|^2 + \fc |\psi_\rc|^2 -1}{D} \,.
\ee
\end{subequations}
For simplicity we focus on purely growing-mode solutions for which the necessary slaving conditions read at $D=0$:
\be
   \fb |\psi_\rb^\ini|^2 + \fc |\psi_\rc^\ini|^2 =1 \,, \qquad   {\cal S}_\alpha^\ini = - \varphi^\ini \,.
\ee
Similarly as in the two-fluid case, the initial component densities must be non-vanishing to avoid quasi-singular behaviour at arbitrary short times; furthermore, from the second boundary condition it is clear that the initial phase for the two fluids coincide. The initial wave function is thus
\be
  \label{eq:psi_ini2fluid}
\psi_\alpha^{\rm ini}(\fett{q}) = \sqrt{1+ \delta_\alpha^{\rm ini}(\fett{q})} \, \exp \left[ \frac{\rm i}{\hbar} \varphi^{\rm ini}(\fett{q}) \right]\,.
\ee
As it was the case for a single fluid, we can use standard PT to obtain solutions to the effective potentials $V_{{\rm eff},\alpha}=3g({\cal S}_\alpha+\varphi)/(2D)$ entering in the Schr\"odinger equations~\eqref{eq:Schroedi2}. 
For the non-decaying mode initial conditions specified here, the velocities of the two fluid species agree at all orders, as we demonstrate in Appendix~\ref{app:2fluid2SPT}. This means that the perturbative effective potentials of the two fluid species are identical with the single fluid effective potential, i.e.,  $V_{{\rm eff},\alpha}^{(n)}\equiv V_{{\rm eff}}^{(n)}$ (cf.\ after Eq.\,\eqref{eq:S2}) and hence
\be 
 \boxed{ 
    \ii \hbar \partial_D \psi_\alpha(\fett{x},D) =  \left[ -\frac{\hbar^2 \nab^2_x}{2}  +  V_{\rm eff}(\fett{x}, D) \right] \psi_\alpha(\fett{x},D) \,. 
 }
\ee

At the leading order, we have $V_{{\rm eff},\alpha}^{(1)}=0$ and thus the solution to the free Schr\"odinger equation for component $\alpha=\rb,\rc$ is given by
\be
\label{eq:PsiKernel_free}
  \psi_{\alpha}^{\rm free}(\fett{x};D) =  \int \! \dd^3 q \, K_\free(\fett{q},\fett{x};D) \,  \psi^{\rm ini}_\alpha(\fett{q}) \,,
\ee
where the free propagator is given by equation~\eqref{eq:K0sol}. Thus, apart from the initial density fluctuations inherent to the component fluids, the solution coincides precisely with the one obtained in the single-fluid case.
Very similarly, the derivations at next-to-leading order are essentially identical as outlined above, with the solution for the wavefunction
\be
\label{eq:PsiKernel_2PPT}
  \psi_{\alpha}^{\text{\scriptsize 2PPT}}(\fett{x};D) =  \int \! \dd^3 q \, K_\twoPPT(\fett{q},\fett{x};D) \,  \psi^{\rm ini}_\alpha(\fett{q}) \,,
\ee
where $K_\twoPPT$ is given by Eq.\,\eqref{eq:KNLOsol} in terms of the second-order effective potential~\eqref{eq:Veff}.

Having obtained the component wavefunction to the desired order in PPT, the corresponding Eulerian density $\rho_\alpha = 1 + \delta_\alpha$ and momentum density $\fett{\pi}_\alpha = \rho_\alpha \fett{v}_\alpha$ are respectively given by
\begin{subequations}
\begin{align}
\rho_\alpha &= \psi_\alpha\,\overline{\psi}_\alpha \,, \\
\fett{\pi}_\alpha&= \frac{{\rm i}\hbar}{2} \left( \psi_\alpha \fett{\nabla} \overline{\psi}_\alpha - \overline{\psi}_\alpha \fett{\nabla} \psi_\alpha \right)\,,
\end{align}
\end{subequations}
where an overline denotes complex conjugation.

Finally we determine  the classical limits of the derived PPT solutions for the component fluids. Actually, 
since the propagator for the component fluids agrees with the one for the single fluid, the classical limit
can be performed precisely with the same methodology as outlined in \S\ \!VI of~\cite{Uhlemann:2019}, however
now generalized to $\Lambda$CDM.
\begin{subequations}
In the classical limit $\hbar \to 0$, we find the following displacement and corresponding velocity valid up to 2PPT
\begin{align}
  &\fett{\xi}_\alpha^\twoPPT =  D \,\fett{\zeta}^{\rM(1)} +  D^2 \fett{\zeta}^{\rM(2)} \,, \\
  &\fett{v}^\twoPPT_\alpha =  \fett{\zeta}^{\rM(1)} +  2D \fett{\zeta}^{\rM(2)}  +  D^2 \zeta^{\rM(1)}_{l} \fett{\zeta}^{\rM(2)}_{,l} \,, \label{eq:v2PPT}
\end{align}
\end{subequations}
where the purely spatial functions $\fett{\zeta}^{\rM(1)}$ and $\fett{\zeta}^{\rM(2)}$ are given in Eqs.\,\eqref{eq:sol2LPTsingle}, 
and, similarly as in the classical case, the corresponding mass conservation law for the above displacement is
$\delta_\alpha = (1+ \delta_\alpha^\ini)/\det[\delta_{ij}+ \xi_{\alpha i,j}^\twoPPT]-1$.
We remark that the ``additional'' term $\sim D^2$ in~\eqref{eq:v2PPT} would be of third order in the classical Lagrangian-coordinates
approach, but here arises naturally in order to preserve the underlying Hamiltonian structure in the present approach. The appearance
of this term was first noted by~\cite{Uhlemann:2019}, where it was also demonstrated that this term is actually needed to preserve the assumed zero-vorticity condition (cf.\ their Fig.\,6).

For numerical applications, keeping $\hbar$ nonzero is crucial as the numerical complexity becomes very demanding in the limit $\hbar \, \dot\to\, 0$, due to strong oscillations of the complex wavefunction.
For nonzero $\hbar$ which effectively controls the resolution in the phase-space, PPT has significant advantages as compared to classical Eulerian perturbative schemes:
On the one hand, the density and velocity fields in PPT are essentially derived by propagating initial fields along the fluid flows -- in a fairly similar way as one follows fluid particles in classical Lagrangian-coordinate approaches. Since Lagrangian perturbative approaches are naturally very efficient in resolving convective motion, roughly the same is true for PPT.
On the other hand,  PPT outputs directly Eulerian fields and thus does not require any $N$-body particle realization which can lead to discretization errors \citep[see e.g.][]{Michaux:2020}. Not relying on particle sampling is a significant advantage particularly for hydrodynamical simulations that usually require Eulerian fields for their initialization. For further details and numerical implementation of PPT, see our companion paper \citep{HahnAccomp2020}.

\section{Summary and outlook} \label{sec:conclusion}

\paragraph*{Summary.} 
Given two gravitationally coupled fluids that are governed by the equations~\eqref{eqs:fluidsD}, 
it becomes evident that relative effects between the fluids are mathematically described {\it by all but the fastest growing modes}. Therefore, as a first approximation, one may include just the strongest of the sub-leading (decaying) modes in the analysis. As it turns out, the most persisting sub-leading mode stems from initially prescribed density perturbations of the two fluids, which are constant in linear theory but nonetheless grow non-linearly in time.

Curiously, there is no zeroth-order approximation in the present case, and we must keep those initial density perturbations in the fluids. Indeed, if we had ignored those initial densities and just kept the very fastest growing modes, a quick analysis would have revealed mathematical inconsistencies that are accompanied with quasi-singular irregularities in the governing equations.

The rigorous argument of the above implies certain boundary conditions on the initial conditions, which are compatible with the requirement of initial quasi-homogeneity. Actually, these boundary conditions are known  but, so far, have been exploited only for the single-matter fluid where they build the mathematical foundation of perturbative solutions in powers of the linear structure growth. In this paper, we have generalized the boundary conditions to allow for multiple fluids (Eqs.\,\ref{eq:slaving2fluid}), thereby providing the stepping stone for initializing two-fluid numerical simulations in the growing and persisting modes which we discuss in detail in our companion paper \citep{HahnAccomp2020}.

Even more, these boundary conditions translate straightforwardly into explicit all-order solutions for the difference density in Eulerian coordinates (Eq.\,\ref{eq:recdeltabcMT}), as well as to the non-linear displacement fields for the two fluids in Lagrangian coordinates. We show that, with a suitable choice of Lagrangian coordinates, the two-fluid displacements (Eq.\,\ref{eq:allorderdispl2f}) actually coincide with the standard ones for the single-matter fluid. Essentially, in those coordinates, the initial fluid densities are just transported along their fluid paths (cf.\ Eq.\,\ref{eq:masscomp}).

Alternatively, one may absorb the initial densities by means of a redefinition of the Lagrangian coordinate system, which however clutters the solutions for the fluid displacements (Eq.\,\ref{eq:comm-prosite-map}). Furthermore, as we elucidate in~\cite{HahnAccomp2020}, absorbing the initial density in the displacements leads to the excitation of large discreteness errors in the numerical solution. We remark that previous approaches for initializing two-fluid  simulations, such as the ones of \cite{Hahn:2011,Valkenburg:2017}, implicitly perform such an operation.

We have also considered a semi-classical approach for two fluids, which largely builds on the work of~\cite{Uhlemann:2019} that we have generalized here to a $\Lambda$CDM cosmology for two fluids. We have motivated the semi-classical approach by a variational principle employing the so-called contact geometry (\S\ref{sec:H&HJE}), which may be viewed as an extension to the symplectic geometry known from standard Hamiltonian theory. In the semi-classical approach, we establish complementary results using the {\it propagator perturbation theory} (PPT) up to second order, which delivers wavefunctions for the two coupled fluids (Eq.\,\ref{eq:PsiKernel_2PPT}) that reproduce in the classical limit the component displacements.

\paragraph*{Outlook.}  There are several avenues that could be considered in future works. In the present study we ignore baryonic pressure, which hampers our theoretical prediction close to the Jeans scale. Incorporating pressure, on the other hand, possibly along the ways of the single-fluid Lagrangian approach of \cite{2002PhRvD..66f4014T} or of  the two-fluid Eulerian approach of \cite{2009ApJ...700..705S}, could comprise promising starting points for initializing $N$-body or hydrodynamical simulations accurately on much smaller scales than anticipated in this work.

Another interesting application of our approach relates to incorporating all decaying modes, which would allow the initialization of several fluids with distinct velocities. This involves in particular the accurate modelling of the advection of small-scale perturbations by large-scale cosmological flows, as has been investigated by~\cite{Tseliakhovich:2010} -- a problem for which our Lagrangian and semi-classical approaches might be ideally suited. Nonetheless, while our general framework is capable of incorporating all decaying modes (\S\ref{sec:generalLag2F}), at this stage it is unclear how respective higher-order initial conditions for simulations could be consistently implemented and thus, such issues require future investigations.

Our two-fluid approach could be also straightforwardly extended to redshift space. Indeed, it is well known that redshift-space distortions are easily incorporated in Lagrangian coordinates \citep[see e.g.][]{Matsubara:2007wj}, and, as shown recently by \cite{Porqueres:2020}, the same is also true for PPT. Once incorporated for multiple fluids in Lagrangian coordinates, our formalism could be used to determine the power spectra in redshift space to arbitrary high order. For the semi-classical approach, a PPT extension to redshift space could provide accurate theoretical predictions that relates the matter distribution to  quasars in the Lyman-$\alpha$ forest, particularly including the correct description of the scale-dependent bias of baryons relative to the total matter distribution. Our work thus provides encouraging starting points to account more accurately for non-linear effects in the fluids, which is essential for  both the forward modelling as well as the reconstruction problem based on observations of the high redshift intergalactic medium.

Finally, while we have focused in this paper on two fluids, generalizations to more fluids are very straightforward. We remark, however, that the ``obvious'' case of three fluids, i.e., including baryons, CDM and massive neutrinos might be best tackled by marrying a two-fluid approach for baryons and CDM with a relativistic description of, e.g., \cite{2017MNRAS.466L..68B,2019JCAP...03..022T,2019MNRAS.489.5938Z,Partmann2020}, that effectively incorporates massive neutrinos (in simulations) by means of suitable coordinate transformations. We will come back to this problem in a  forthcoming work.

\section*{Acknowledgements}

We thank Vincent Desjacques, Uriel Frisch, and Andrei Sobolevski\u{\i} for useful discussions. 
C.R.\ is a Marie Sk\l odowska-Curie Fellow and acknowledges funding from the People Programme (Marie Curie Actions) 
of the European Union's Horizon 2020 Programme under Grant Agreement No.\ 795707 (COSMO-BLOW-UP). 
 O.H.\ acknowledges funding from the European Research Council 
(ERC) under the European Union's Horizon 2020 research and innovation programme, 
Grant agreement No.\ 679145 (COSMO-SIMS).

\section*{Data Availability}

There are no new data associated with this article.



\bibliographystyle{mnras}
\bibliography{bibliography}


\appendix

\section{Single-fluid solutions in weak expansion}\label{app:1fluid2LPT}

For completeness, here we review the standard results for a single fluid in $\Lambda$CDM, assuming a weak perturbation expansion of the form
\be
  \delta_\rM = \epsilon \,\delta_\rM^{(1)} + \epsilon^2 \,\delta_\rM^{(2)}+ \ldots \,.
\ee
Naturally one {\it could} assume that each order $\delta_\rM^{(1)}$ factorizes into a purely space and time dependent part, at least for the fastest growing-mode solutions. 
However, \cite{Bouchet:1994xp,Matsubara:1995kq} showed that this is not the case for the weak expansion of the Eulerian density in $\Lambda$CDM; see \cite{Villa:2015ppa} for a more recent discussion in a relativistic context. 

To make progress, even for determining the Eulerian density, it is -- not necessary but advantageous -- to solve the problem firstly in Lagrangian space, and then subsequently transform those results to Eulerian space. To do so, we solve the Lagrangian evolution equations~\eqref{eq:sol2LPTsingle} with the following weak expansion {\it Ansatz} for the displacement (the dynamical quantity in Lagrangian space),
\begin{align}
  \fett{\xi}_\rM &= \epsilon \,\bar D(t) \, {\fett{\xi}}^{\rM (1)}(\fett{q})  + \epsilon^2 \,\bar E(t) \,{\fett{\xi}}^{\rM (2)}(\fett{q})+ \ldots  \,,
\end{align}
where we have assumed (correctly) that the Lagrangian solutions factorize, and we have added the bar on top of some quantities to distinguish between the different expansion schemes used in this paper. It is also useful to provide the corresponding Jacobian up to second order
\be
  J = \det[\delta_{ij}+ \xi_{i,j}^\rM] = 1 + \epsilon \bar D \, \xi_{l,l}^{\rM (1)} + \epsilon^2 \bar E \,\xi_{l,l}^{\rM (2)} + \epsilon^2 \bar D^2  \mu_2^{\rM (1,1)} ,
\ee
where $\mu_2^{\rM (1,1)}=(1/2)[\xi_{i,i}^{\rM (1)} \xi_{j,j}^{\rM (1)} - \xi_{i,j}^{\rM (1)} \xi_{j,i}^{\rM (1)}]$.

At first order in~$\epsilon$, the space part  is determined by using the boundary conditions (see \S\ref{sec:singlefluidLPT}),
leading to ${\fett{\xi}}^{\rM (1)} = -\nab \varphi^{\rm ini}$, 
 while one gets an ODE for the temporal coefficient
\be \label{eq:Dbar}
  \RT \bar D - \frac{3g}{2D^2} \bar D= 0  \,,
\ee
where as before $\RT = (\partial_D^\rL)^2 +[3 g/(2D)] \partial_D^\rL$.
Due to the appearance of the time-dependent factor $g(D)$, this ODE is most easily solved numerically. We show the numerical solution in Fig.\,\ref{fig:DandE}, which agrees with the standard analytical solution~\eqref{eq:Dplus} to machine precision, thus suggesting that effectively $g=1$ at this order, and thus we set from here on  $\bar D = D$.

To get the second-order solution, we truncate all terms $O(\epsilon^2)$ in~\eqref{eq:sol2LPTsingle} and get for the spatial part $\xi_{l,l}^{\rM (2)} = \mu_2^{\rM (1,1)}$ which again coincides with the standard result. By contrast, for the temporal coefficient we obtain the ODE 
\be \label{eq:barE}
  \RT \bar E - \frac{3g}{2D^2} \bar E = -\frac{3g} 2 \,.
\ee
While in EdS, this equation can still be solved analytically ($\to \bar E \sim (-3/7)a^2$), there is no known analytical solution in $\Lambda$CDM -- as opposed to the derived analytical solution derived in terms of the $D$ expansion where $\bar E \to (-3/7)D^2$. However, the numerical solution for~\eqref{eq:barE} is displayed in Fig.\,\ref{fig:DandE}, and the comparison against $\bar E \to (-3/7)D^2$ reveals excellent agreement. Therefore we set in the following $\bar D = D$ and $\bar E = E =  (-3/7)D^2$.

To get an expression for the density in terms of the Lagrangian solution, we Taylor expand the Lagrangian mass conservation $\delta=1/J-1$ to second order, yielding firstly
\be
  \delta_\rM(\fett{q}) = - D \varphi_{,ll}^{\rm ini} +  \frac{D^2- E}{2} \varphi_{,ll}^{\rm ini} \varphi_{,mm}^{\rm ini}  + \frac{D^2 + E}{2} \varphi_{,lm}^{\rm ini} \varphi_{,lm}^{\rm ini} \,.
\ee
Finally, to obtain the density evaluated at the current (Eulerian) position, we use the inverse map $\fett{q}(\fett{x}) = \fett{x}-\fett{\xi}$ to first order in
$\delta_\rM(\fett{q}(\fett{x}))=\delta_\rM(\fett{x}-\fett{\xi})$. As a consequence, the transported first-order solution generates a second-order term (a.k.a.\ the convective term), leading to the second-order result for the Eulerian density
\be \label{eq:d2LCDM}
  \delta_\rM^{(2)}(\fett{x}) = \frac{D^2-E}{2} \varphi_{,ll}^{\rm ini} \varphi_{,mm}^{\rm ini} \!+  D^2 \varphi_{,llm}^{\rm ini} \varphi_{,m}^{\rm ini} + \frac{D^2 +E}{2} \varphi_{,lm}^{\rm ini} \varphi_{,lm}^{\rm ini} \,,
\ee
which agrees with the reported result given in Eq.\,\eqref{eq:soldelta2weak}.
As promised above, from the solution~\eqref{eq:d2LCDM} it is clear that, in Eulerian
coordinates, spatial and temporal dependencies do not factorize at second order in~$\Lambda$CDM, a fact that was first noted by~\cite{Matsubara:1995kq}.

\vskip-0.6cm

\phantom{1}

\section{Two-fluid solutions in second-order perturbation theory}\label{app:2fluid2SPT}

\subsection{Two-fluids assuming single-fluid Poisson source}

Set $\delta_\alpha = \delta_\alpha^{(1)} + \delta_\alpha^{(2)}$ and $\phi_\alpha = \phi_\alpha^{(1)} + \phi_\alpha^{(2)}$, where $\fett{v}_\alpha \equiv - \nab \phi_\alpha$, for $\alpha=b,c$.
The first-order growing-mode solutions derived in the main text are
\begin{align} \label{eq:soldeltaini}
 \begin{aligned}
  \delta_\alpha^{(1)} &= D \nab^2 \varphi^{\rm ini} + \delta_\alpha^{\rm ini} \,, \\
  \phi_\alpha^{(1)} &=  \varphi^{\rm ini} \,,
 \end{aligned}
\end{align}
where 
\be
   \delta_\alpha^{\rm ini} = \left\{
    \begin{matrix}
      (1- \fb)\, \delta_\bc^\ini \,,  &\alpha = \rb \,, \\
        - \fb   \delta_\bc^\ini \,,    &\alpha = \rc \,.
    \end{matrix} \right.  
\ee
 The non-linear fluid equations for the components $\alpha=b,c$ can be written as
\begin{align}
 &\partial_D \phi_\alpha - \frac 1 2 |\nab \phi_\alpha|^2 = \frac{3g}{2D} \left( \varphi - \phi_\alpha \right) \,, \\
 &\partial_D \delta_\alpha - \nab \cdot (1+\delta_\alpha ) \nab \phi_\alpha = 0 \,,  \\
 &\nab^2 \varphi = \frac {\delta_{\rm m}} D \,.
\end{align}
At second order we have $g\simeq 1$, and thus these equations become
\begin{align}
 &\partial_D \phi_\alpha^{(2)} - \frac 1 2 |\nab \phi_\alpha^{(1)}|^2 = \frac{3}{2D} \left( \varphi^{(2)} - \phi_\alpha^{(2)} \right) \label{eulereA2}\,, \\
 & \partial_D \delta_\alpha^{(2)} - \nab^2 \phi_\alpha^{(2)} - \nab (\delta_\alpha^{(1)} \nab \phi_\alpha^{(1)}) = 0 \,, \label{contiA2} \\
 &\nab^2 \varphi^{(2)} = \frac {\delta_{\rm m}^{(2)}(\fett{x},D)} D \,.
\end{align}
In these equations all quantities with a perturbation index of~1 are already determined; furthermore we have already 
derived $\delta_{\rm m}^{(2)}(\fett{x},D) = D^2 \delta_\rM^{(2)}(\fett{x})$ in the main text, see Eq.\,\eqref{eq:2SPT}, implying that we also know already $\varphi^{(2)}$.
Rewriting~\eqref{contiA2} in terms of $\phi_\alpha^{(2)}$ and plugging it into~\eqref{eulereA2}, we get
\begin{align}
  &\partial_D^2 \delta_\alpha^{(2)} + \frac{3}{2D} \partial_D \delta_\alpha^{(2)} 
 = 5 \delta_\rM^{(2)}(\fett{x}) + \frac{3}{2D}  \left[ \delta_\alpha^{\rm ini} \varphi_{,mm}^{\rm ini} +  \delta_{\alpha,m}^{\rm ini} \varphi_{,m}^{\rm ini}\right] \,.
\end{align}
The general solution to this equation is
\be \label{eq:soldeltaA}
  \delta_\alpha^{(2)} = D^2 \delta_\rM^{(2)}
    - 2 C_1 D^{-1/2} + C_2  + D \left[ \delta_\alpha^{\rm ini} \varphi_{,mm}^{\rm ini} +  \delta_{\alpha,m}^{\rm ini} \varphi_{,m}^{\rm ini}\right] \,,
\ee
where $C_1$ and $C_2$ are integration constants; actually the term involving $C_1$ is a decaying mode and not compatible with slaving, thus $C_1 =0$. The other term, by contrast, is physically redundant as it can be absorbed into $\delta_\bc^{\rm ini}$, thus we can set $C_2=0$.
Plugging the solution for $\delta_2^\alpha$ into~\eqref{contiA2} we then find the solution for the second-order velocity potential, i.e., 
\be
  \nab^2 \phi_\alpha^{(2)} /D= \frac 3 7 \varphi_{,ll}^{\rm ini} \varphi_{,mm}^{\rm ini} + \varphi_{,llm}^{\rm ini} \varphi_{,m}^{\rm ini} + \frac 4 7 \varphi_{,lm}^{\rm ini} \varphi_{,lm}^{\rm ini} \,,
\ee
where, evidently, all terms involving $\delta_\bc^{\rm ini}$ have cancelled out. Finally, using these results, we find that the so-called effective potential for the components $\alpha=b,c$ is
\be \label{eq:Veffapp}
  V_{{\rm eff}, \alpha} \!= \!\frac{3g}{2D} \left( \varphi - \phi_\alpha \right) = \frac 3 7 \nab^{-2}\! \left[  \varphi_{,ll}^{\rm ini} \varphi_{,mm}^{\rm ini} -  \varphi_{,lm}^{\rm ini} \varphi_{,lm}^{\rm ini}  \right] \! + O(3).
\ee
This effective potential is the necessary input to determine the NLO propagator in PPT.

\vskip-0.7cm

\phantom{1}

\subsection{Growing-mode solutions for sum and difference variables}

Here we are concerned with deriving the second-order solutions by employing the weighted sum and difference variables 
\begin{align}
 \begin{aligned}
   \delta_\rM &= \fb \delta_\rb + \fc \delta_\rc  \,,  \qquad &\delta_\bc &=  \delta_\rb -  \delta_\rc \,,  \\
   \fett{v}_\rM &= \fb \fett{v}_\rb + \fc \fett{v}_\rc  \,, \qquad  &\fett{v}_\bc  &=  \fett{v}_\rb -  \fett{v}_\rc 
 \end{aligned}
\end{align}
at first and second order in PT, i.e., 
\begin{align}
 \begin{aligned}
  \delta_\rM &= \delta_\rM^{(1)} + \delta_\rM^{(2)} + \ldots \,,  \qquad &\delta_\bc &= \delta_\bc^{(1)} + \delta_\bc^{(2)} + \ldots  \,,  \\
  \fett{v}_\rM &=  \fett{v}_\rM^{(1)} + \fett{v}_\rM^{(2)} + \ldots \,,   \qquad  &\fett{v}_\bc  &=  \fett{v}_\bc^{(1)} + \fett{v}_\bc^{(2)} + \ldots \,, \\
   \delta_\rb &= \delta_\rb^{(1)} + \delta_\rb^{(2)} + \ldots \,,  \qquad &\delta_\rc &= \delta_\rc^{(1)} + \delta_\rc^{(2)} + \ldots  \,,  \\
  \fett{v}_\rb &=  \fett{v}_\rb^{(1)} + \fett{v}_\rb^{(2)} + \ldots \,,   \qquad  &\fett{v}_\rc  &=  \fett{v}_\rc^{(1)} + \fett{v}_\rc^{(2)} + \ldots \,,  
 \end{aligned}
\end{align}
From the main text we have already derived the growing-mode solutions at first order, which we summarize here for convenience
\begin{align}
 \begin{aligned}
  \delta_\rM^{(1)} &= D\, \nab^2 \varphi^{\rm ini} \,,  \qquad &\delta_\bc^{(1)} &= \delta_\bc^\ini \,, \\
  \fett{v}_\rM^{(1)} &= - \nab \varphi^{\rm ini} \,,  \qquad &\fett{v}_\bc^{(1)} &= 0 \,, \\
   \delta_\rb^{(1)} &= \delta_\rM^{(1)} +\delta_\rb^{\rm ini} \,, \qquad  &\delta_\rc^{(1)} &= \delta_\rM^{(1)}+\delta_\rc^{\rm ini} \,, \\
  \fett{v}_\rb^{(1)} &= \fett{v}_\rM^{(1)} \,,  \qquad &\fett{v}_\rc^{(1)} &= \fett{v}_\rM^{(1)} \,, \\
 \end{aligned}
\end{align}
where $\delta_\rb^{\rm ini} =  (1-\fb) \, \delta_\bc^\ini$ and $\delta_\rc^{\rm ini} =  -\fb \, \delta_\bc^\ini$. As before these growing-mode results employ implicitly the slaving conditions~\eqref{eq:slaving2fluid}.
Using these results in the component fluid equations~\eqref{eqs:fluidsD}, 
we have at second order for the components
\begin{subequations} \label{eqs:fluidsDapp}
 \begin{align}
& \partial_D \fett{v}_\alpha^{(2)} + \fett{v}_\rM^{(1)} \cdot \nab \fett{v}_\rM^{(1)} = - \frac{3g}{2D} \big( \fett{v}_\alpha^{(2)} + \nab \varphi^{(2)} \big) \,,   \label{eq:compeuler2nd}   \\
& \partial_D \delta_\alpha^{(2)} + \nab \cdot \fett{v}_\alpha^{(2)} + \nab \cdot [\delta_\rM^{(1)}\, \fett{v}_\rM^{(1)}]  + \nab \cdot [\delta_\alpha^{\rm ini} \,\fett{v}_\rM^{(1)}] =0 \,,  \label{eq:compconti2nd}  \\
&\nab^2  \varphi^{(2)} =  \delta_\rM^{(2)}/ D \,,   
 \end{align}
\end{subequations}
where, for convenience, we have already expressed some first-order component variables in terms of the single fluid variables.
These equations can be written in terms of the weighted sum and difference variables, we find truncated up to second order for the growing modes
\begin{subequations} \label{eqs:fluidsDapp2}
 \begin{align}
& \partial_D \fett{v}_\rM + \fett{v}_\rM \cdot \nab \fett{v}_\rM +O(3) = - \frac{3g}{2D} \big( \fett{v}_\rM + \nab \varphi \big)  \,,   \label{eq:compeuler2ndmatter}   \\
& \partial_D \delta_\rM + \nab \cdot ([1+\delta_\rM] \,\fett{v}_\rM)  + O(3) = 0 \,,   \label{eq:compconti2ndmatter}  \\
& \partial_D \fett{v}_\bc  + \frac{3g}{2D} \fett{v}_\bc  + O(3) =0  \,,   \label{eq:compeuler2nddiff}   \\
& \partial_D \delta_\bc + \nab \cdot  \fett{v}_\bc  - \nab \cdot [ \delta_\bc^\ini\, \nab \varphi^{\rm ini}]  + O(3) = 0 \,.   \label{eq:compconti2nddiff} 
 \end{align}
\end{subequations}
To arrive at~\eqref{eq:compconti2ndmatter} we have explicitly used the slaving condition $\delta_\rM^{\rm ini}=0$. Crucially, the sum and difference equations still decouple effectively at second order. While the truncated sum equations are identical with the standard equations for a single fluid, the continuity equation for the difference variables receives a second-order correction. Specifically, combining~\eqref{eq:compeuler2nddiff}--\eqref{eq:compconti2nddiff} we obtain at second order the ODE
\be
  \delta_D^2 \delta^{(2)}_\bc + \frac{3g}{2D}  \delta_D \delta^{(2)}_\bc  = \frac{3g}{2D} \nab \cdot (\delta_\bc^\ini \nab \varphi^\ini) \,,
\ee
which for $g\simeq 1$ has the analytic solution
\be
  \delta^{(2)}_\bc =  D \, \nab \cdot (\delta_\bc^\ini \nab \varphi^\ini) - \frac{2 C_1}{\sqrt{D}} + C_2 \,,
\ee
where $C_1 \to 0$ due to slaving, and $C_2$ can be set to zero (or, equivalently, absorbed in the first-order constant).
It is easily checked that having the sum and difference solutions, one re-derives the identical solution~\eqref{eq:soldeltaA} from $\delta_\rb^{(2)} = \delta_\rM^{(2)} + (1- \fb)\, \delta_\bc^{(2)}$ and $\delta_\rc^{(2)} = \delta_\rM^{(2)}  - \fb\, \delta_\bc^{(2)}$.

\subsection{All-order recursions for the difference density}\label{app:sumdifference-nonpert}

From the above considerations it is clear that for growing-mode initial conditions we have at least to second order $\fett{v}_\bc =0$ and thus the component velocities coincide. Actually, using this and iterating the expressions~\eqref{eqs:fluidsDapp2} to third order, one finds that $\fett{v}_\bc =0$  is also satisfied at third order. 
Further iterations then lead to the conclusions that $\fett{v}_\bc$ must be zero to all orders in PT, essentially since all appearing terms at higher orders in~\eqref{eq:compeuler2ndmatter} are quadratic combinations of the lower-order velocity components stemming from convective terms such as $\fett{v}_\alpha \cdot \nab \fett{v}_\alpha$; but since those lower-order velocity components achieve $\fett{v}_\bc =0$, this also implies that $\fett{v}_\rb= \fett{v}_\rc = \fett{v}_\rM$ to all orders. Thus, the convective term in the evolution equation for $\fett{v}_\bc$ drops out to all orders.

In summary the sum and difference fluid equations are nonperturbatively for the growing mode 
\begin{subequations} \label{eqs:fluidsDapp3}
 \begin{align}
& \partial_D \fett{v}_\rM + \fett{v}_\rM \cdot \nab \fett{v}_\rM  = - \frac{3g}{2D} \big( \fett{v}_\rM + \nab \varphi \big)  \,,   \label{eq:compeuler2ndmatterNP}   \\
& \partial_D \delta_\rM + \nab \cdot ([1+\delta_\rM] \,\fett{v}_\rM) = 0 \,,   \label{eq:compconti2ndmatterNP}  \\
& \partial_D \fett{v}_\bc  + \frac{3g}{2D} \fett{v}_\bc  =0  \,,   \label{eq:compeuler2nddiffNP}   \\
& \partial_D \delta_\bc + \nab \cdot \left[ \delta_\bc \fett{v}_\rM \right] = 0 \,.   \label{eq:compconti2nddiffNP} 
 \end{align}
\end{subequations}
Two remarks are in order. First, as mentioned above, from~\eqref{eq:compeuler2nddiffNP} follows that the growing-modes of $\fett{v}_\bc$ must be zero at all orders. Secondly, 
since $\fett{v}_\rM = \fett{v}_\alpha$ nonperturbatively, from the mass conservation~\eqref{eq:compconti2nddiffNP} it is clear that $\delta_\bc$  couples only to the sum velocity $\fett{v}_\rM$. For the latter there exist explicit all-order recursion relations for the growing mode, they can be written as 
\be
  \theta_\rM  = - \sum_{n=1}^\infty \theta_\rM^{(n)} D^{n-1}   = - \sum_{n=1}^\infty \nab \cdot \fett{v}_\rM^{(n)} D^{n-1} = \nab \cdot \fett{v}_\rM \,,
\ee 
where $\theta_\rM^{(n)}$ is the $n$th-order perturbation kernel which in the literature are usually formulated in Fourier space and then denoted with $G_n$(see however Eq.\,(10) of~\citealt{Taruya:2018} for a real-space version).
From this it is easily checked that
the appropriate {\it Ansatz} for $\delta_\bc$ is
\be
  \delta_\bc = \sum_{n=1}^\infty \delta_\bc^{(n)} D^{n-1} \,.
\ee
Plugging the {\it Ans\"atze} for $\delta_\bc$ and $\fett{v}_\rM$ into~\eqref{eq:compconti2nddiffNP} and using that $\theta_\bc =0$, we obtain the following all-order recursion relation
\be \label{eq:recdeltabc}
   \delta_\bc^{(n)} = \frac{1}{n-1} \sum_{i+j=n} \nab \cdot \left[ \delta_\bc^{(i)} \fett{v}_\rM^{(j)} \right] 
\ee
for $n>1$, and $\delta_\bc^{(1)}= \delta_\bc^\ini$ for $n=1$.

We remark that a similar relation for $\delta_\bc$ can also be formulated in Lagrangian space:
Writing $\delta^\alpha (\fett{x}(\fett{q})) = (1+ \delta_\alpha^\ini (\fett{q}))/ J^\alpha(\fett{q}) - 1$ and noting that for the growing modes we have $J^\alpha = J^\rM$, we can obtain the difference density by simply subtracting the two definitions of mass conservation, i.e.,
\begin{align}
\boxed{
 \begin{aligned}
    &\delta^\bc (\fett{x}(\fett{q})) = \delta^\rb (\fett{x}(\fett{q})) - \delta^\rc (\fett{x}(\fett{q}))
   = \frac{\delta_\bc^\ini(\fett{q})}{J^\rM(\fett{q})}  \\  
   &\,\,\,= \delta_\bc^\ini(\fett{q}) \Big[ ( 1 + D \varphi_{,ll}^\ini) + \left( \frac 5 7 \varphi_{,ll}^2  + \frac 2 7 \varphi_{,lm}^2  \right) D^2 \Big] + O(4) \,,
 \end{aligned}
}
\end{align}
which, when evaluated at the Eulerian position to fixed order, delivers identical results as from~\eqref{eq:recdeltabc}. We remark that to verify the agreement to order $n=3$, one needs to evaluate the fields at the position $q_i(\fett{x}) = x_i - \xi_i^{\rM(1)}(\fett{x}) - \xi_i^{\rM(2)} + \xi_{i|l}^{\rM(1)} \xi_l^{\rM(1)}$, where the slash denotes differentiation with respect to Eulerian coordinates.

Finally, let us provide these recursion relations in Fourier space, we find
\begin{align} \label{eq:deltabcFourier}
  \tilde \delta_\bc^{(n)} (\fett{k}) &= \int \frac{\dd^3 k_1 \cdots \dd^3 k_n}{(2\pi)^{3n}}
    \delta_{\rm D}^{(3)} (\fett{k}_{1 \cdots n} - \fett{k}) F_\bc^{(n)} (\fett{k}_1, \ldots, \fett{k}_n) \nonumber \\ 
    &\qquad  \times \tilde \delta^{(1)}_\bc(\fett{k}_1) \,\tilde \delta_\rM^{(1)}(\fett{k}_2) \cdots \tilde \delta_\rM^{(1)}(\fett{k}_n) \,, 
\end{align}
where $\fett{k}_{12 \cdots n} = \fett{k}_1 + \fett{k}_2+  \cdots + \fett{k}_n$, and the first kernels are 
\begin{align}
  F^{(1)}_\bc &= 1 \,, \\
  F^{(2)}_\bc &= \frac{\fett{k}_{12} \cdot \fett{k}_2}{k_2^2} \,, \\
  F^{(3)}_\bc &= \frac 1 2  \frac{\fett{k}_{123} \cdot \fett{k}_{23}}{k_{23}^2} \left[  \frac 3 7 +  \frac{\fett{k}_{2} \cdot \fett{k}_3}{2 k_2 k_3} \left( \frac{k_2}{k_3} + \frac{k_3}{k_2} \right)  + \frac 4 7\frac{(\fett{k}_2 \cdot \fett{k}_3)^2}{k_2^2 k_3^2} \right]  \nonumber \\ 
  &\qquad + \frac 1 2\frac{\fett{k}_{12} \cdot \fett{k}_2}{k_2^2}  \frac{\fett{k}_{123} \cdot \fett{k}_{3}}{k_{3}^2} \,,
\end{align}
where $k = |\fett{k}|$.
These kernels are to be symmetrized in their arguments $\fett{k}_2$ -- $\fett{k}_n$, but not in its first argument $\fett{k}_1$; this is a consequence of the Fourier kernels~\eqref{eq:deltabcFourier} within the integrals.
Fairly similar to the well-known density and velocity kernels in SPT \citep[see e.g.][]{Bernardeau:2002}, the kernels $F^{(n)}_\bc$ are well behaved when the sum of some of its arguments cancel, but there are (the known) infrared divergences when one or more of its arguments go to zero. Furthermore and in contrast to the standard SPT kernels,  for $\fett{k}= \fett{k}_{12 \cdots n} =0$, the kernels $F^{(n)}_\bc$ do not asymptote $k^2$ but vanish instead; this so due to the appearance of the overall divergence in the recursion relation~\eqref{eq:recdeltabc}.

\vskip-0.6cm

\phantom{1}

\subsection{One-loop power spectrum for the density difference}\label{app:2fluid-oneloop}

Define the linear power and cross spectra with
\begin{subequations}
\begin{align}
 & \left\langle \tilde \delta_{\rm m}^{(1)}(\fett{k}_1) \,\tilde \delta_{\rm m}^{(1)}(\fett{k}_2 ) \right\rangle = (2\pi)^3 \delta_{\rm D}^{(3)}(\fett{k}_{12}) \,P_{\rm m, m}^{\rm lin}(k_1)  \,, \\
 & \left\langle \tilde \delta_{\rm bc}^{(1)}(\fett{k}_1) \,\tilde \delta_{\rm m}^{(1)}(\fett{k}_2 ) \right\rangle = (2\pi)^3 \delta_{\rm D}^{(3)}(\fett{k}_{12}) \, P_{\rm bc, m}^{\rm lin}(k_1)  \,, \\
 & \left\langle \tilde \delta_{\rm bc}^{(1)}(\fett{k}_1) \, \tilde \delta_{\rm bc}^{(1)}(\fett{k}_2 ) \right\rangle = (2\pi)^3 \delta_{\rm D}^{(3)}(\fett{k}_{12}) \, P_{\rm bc, bc}^{\rm lin}(k_1)  \,. 
\end{align}
\end{subequations}
Here we like to determine the power spectrum for $P_{\rm bc, bc}$ to one-loop accuracy, i.e., approximate
$\delta_\bc = \delta^{(1)}_\bc + \delta^{(2)}_\bc + \delta^{(3)}_\bc$ and derive 
\begin{align}
  P_{\bc,\bc}(k) = P_{\bc,\bc}^{\rm lin}(k) + P_{\bc,\bc}^{\rm one-loop}(k)
\end{align}
with
\be
   P_{\bc,\bc}^{\rm one-loop}(k) = P_{\bc,\bc}^{(2,2)}(k) + 2 P_{\bc,\bc}^{(1,3)}(k) 
\ee
and
\begin{align}
 & \left\langle \tilde \delta_{\bc}^{(1)}(\fett{k}_1) \, \tilde \delta_{\rm bc}^{(3)}(\fett{k}_2 ) \right\rangle = (2\pi)^3 \delta_{\rm D}^{(3)}(\fett{k}_{12}) \, P_{\rm bc, bc}^{(1,3)}(k_1)  \,, \\
 & \left\langle \tilde \delta_{\bc}^{(2)}(\fett{k}_1) \, \tilde \delta_{\rm bc}^{(2)}(\fett{k}_2 ) \right\rangle = (2\pi)^3 \delta_{\rm D}^{(3)}(\fett{k}_{12}) \, P_{\rm bc, bc}^{(2,2)}(k_1)  \,.
\end{align}
Having the explicit expressions for $\tilde \delta_\bc^{(n)}$, it is straightforward to determine these one-loop corrections by applying Wick's theorem \citep[for similar derivation for the matter power spectrum, see e.g.][]{2006PhRvD..73f3519C}. We find the connected parts
\begin{align}
  &P_{\rm bc, bc}^{(1,3)}(k) = P_{\bc, \bc}^{\rm lin}(k) \int \frac{\dd^3 p}{(2\pi)^3} F_\bc^{(3)}(\fett{k}, \fett{p}, -\fett{p})\, P_{\rM, \rM}^{\rm lin}(p) \nonumber \\
 &\qquad \quad\,\,  + 2 P_{\bc, \rM}^{\rm lin}(k) \int \frac{\dd^3 p}{(2\pi)^3} F_\bc^{(3)}(\fett{p}, -\fett{p}, \fett{k})\, P_{\bc, \rM}^{\rm lin}(p) \,, \\
 &P_{\rm bc, bc}^{(2,2)}(k) =  \int \frac{\dd^3 p}{(2\pi)^3} \left( F_\bc^{(2)}( \fett{k} -\fett{p}, \fett{p}) \right)^2 P_{\bc}^{\rm lin}(|\fett{k}- \fett{p}|) P_{\rM}^{\rm lin}(p) \nonumber  \\  & +
 \!\int \!\! \frac{\dd^3 p}{(2\pi)^3} F_\bc^{(2)}(\fett{p}, \fett{k} -\fett{p}) F_\bc^{(2)}( \fett{k} -\fett{p}, \fett{p})  P_{\bc, \rM}^{\rm lin}(|\fett{k}-\fett{p}|)\, P_{\bc, \rM}^{\rm lin}(p) \,.
\end{align}

\section{Two-fluid solutions in Lagrangian perturbation theory}

\subsection{Explicit derivations in the general formalism}\label{app:2fluid2LPT}

In this Appendix we provide the general second-order perturbation equations for the 2-fluid system in Lagrangian coordinates. 
For this we assume a weak expansion in the displacement according to
\be
 \fett{x}^\alpha - \fett{q}=  \fett{\xi}^\alpha =  \fett{\xi}^{(1)\alpha}+ \fett{\xi}^{\alpha(2)} + \ldots \,,
\ee
and assume, as before, that $\delta_\alpha^{\rm ini}$ is not larger than typical first-order perturbations.
Let us define
\begin{align}
  \mu_2^{\alpha,\beta} &= \frac 1 2 \left[ \xi_{i,i}^\alpha \xi_{j,j}^\beta - \xi_{i,j}^\alpha \xi_{j,i}^\beta \right] \\
    &=  \frac 1 2 \left[ \xi_{i,i}^{(1)\alpha} \xi_{j,j}^{(1)\beta} - \xi_{i,j}^{(1)\alpha} \xi_{j,i}^{(1)\beta} \right] + O(3) \,, \nonumber
\end{align}
where $\alpha$ and $\beta$ are either b or c, and
provide some of the related expressions that are needed later on, valid until second order,
\begin{align}
  &J^\alpha(\fett{q})  = 1 + \xi_{l,l}^{(1)\alpha}(\fett{q}) + \xi_{l,l}^{(2)\alpha} + \mu_2^{\alpha,\alpha}  \,, \label{eqJalp} \\
  &x_i^\alpha(q_l) = q_i + \xi_i^{\rm (1)\alpha}(q_l) +  \xi_i^{\rm (2)\alpha}(q_l)  \,, \\
  &q_i^\alpha(x_l) = x_i + \xi_{i|l}^\alpha \xi_l^\alpha - \xi_i^\alpha(x_l)  \,, \label{eq:inversemap} \\
  &q_i^{\rm c}(x_k^{\rm b}) =  q_i + \xi_{i|l}^{\rm (1)c}\left( \xi_l^{\rm (1)c} -  \xi_l^{\rm (1)b}   \right)  - \xi_i^{\rm (1)c} - \xi_i^{\rm c(2)} + 
    \xi_i^{\rm (1)b} +  \xi_i^{\rm b(2)}, \\
 &J^{\rm c}|_{\fett{q}=\fett{q}^{\rm c}(\fett{x}^{\rm b})} =  1 + \xi_{i,i}^{\rm (1)c}  + \xi_{i,ij}^{\rm (1)c} \left(  \xi_{j}^{\rm (1)b} \!-  \xi_{j}^{\rm (1)c} \right)  + \xi_{i,i}^{\rm c(2)} + \mu_2^{\rc,\rc} , \label{eqJc}
\end{align}
and similarly for the nontrivial terms appearing in the evolution equation~\eqref{eq:LEOM1} of the c component. Here, ``$F_{|i}$'' denotes partial differentiation of an arbitrary function $F$ with respect to Eulerian component~$x_i$.

Using these identities, it is straightforward to determine the second-order part of Eq.\,\eqref{eq:LEOM1}, we find
\begin{subequations} \label{eq:EOM2lpt2lfuidsimpleapp}
\begin{align}
   \RT   \xi_{l,l}^{\rm b(2)} \!& =\! \frac{3g}{2D^2} \Bigg[  \fb \! \left( \xi_{l,l}^{\rm b(2)} - \mu_2^{\rb,\rb} \right) +\fc \left( \xi_{l,l}^{\rm c(2)} +  \mu_2^{\rc,\rc} - 2 \mu_2^{\rm b,c}  \right) \nonumber \\ 
     & -\fc \bigg(  (\delta_{\rm c}^{\rm ini} - \xi_{i,i}^{\rm c(1)}) \partial_j  - \xi_{i,ij}^{\rm c(1)}  \bigg)  \bigg\{ \xi_{j}^{\rm b(1)} - \xi_{j}^{\rm c(1)} \bigg\}
\Bigg] \,, 
   \label{eq:EOM2lpt2lfuid1simplifiedapp} \\ 
\RT   \xi_{l,l}^{\rm c(2)} \!&=\! \frac{3g}{2D^2} \Bigg[ \fc \!\left( \xi_{l,l}^{\rm c(2)} - \mu_2^{\rc,\rc} \right)  + \fb \!\left( \xi_{l,l}^{\rm b(2)} + \mu_2^{\rb,\rb} - 2\mu_2^{\rm b,c} \right) \nonumber \\
   & + \fb  \bigg(   (\delta_\rb^{\rm ini} - \xi_{i,i}^{\rm b(1)}) \partial_j - \xi_{i,ij}^{\rm b(1)} \bigg)  \bigg\{ \xi_{j}^{\rm b(1)} - \xi_{j}^{\rm c(1)} \bigg\} \Bigg] \,, 
   \label{eq:EOM2lpt2lfuid2simplifiedapp}
\end{align}
\end{subequations}
In deriving~\eqref{eq:EOM2lpt2lfuid1simplifiedapp}, we have simplified first-order expressions of the kind $\RT   \xi_{l,l}^{\rm (1)\alpha}$ and $\RT   \xi_{l,m}^{\rm (1)\alpha}$ with their respective right-hand side's, as instructed through Eq.\,\eqref{eq:1LPT2fluid} (here occurring integration constants can be safely ignored).

\subsection{Component velocity in the Lagrangian approaches}\label{app:2fluid2LPTvelocity}

In the main text we have derived the component displacement in two complementary approaches. 
Although these displacements appear substantially different,
we have already shown in the main text that both displacement achieve the same Eulerian density, which is an important consistency check. Here we show that the same is also true for the Eulerian velocity.

For both Lagrangian-coordinate approaches, it is useful to determine the Eulerian velocity by considering the Eulerian continuity equation for the components given in Eq.\,\eqref{eq:massDalpha}, which can equivalently be written as
\be
 \partial_D^{\rm L} \delta_\alpha + (1+ \delta_\alpha ) \,\theta_\alpha =0 \,, 
\ee
where $\theta_\alpha = \nab_x \cdot \fett{v}_\alpha$, and
 $\partial_D^{\rm L}$ denotes, as before, the convective time derivative. Since we have already derived the density in both Lagrangian approaches parametrized through the Lagrangian coordinate $\fett{q}$, we use the above equation to determine the corresponding velocity divergence.
For this we consider a weak expansion of all involved fields of the form
\be
  \delta_\alpha = \delta_\alpha^{(1)} + \delta_\alpha^{(2)} \,, \qquad \theta_\alpha = -\theta_\alpha^{(1)} - \theta_\alpha^{(2)} 
\ee 
 (note the  minus sign in the second {\it Ansatz} due to convention),
which leads to the following constraint equations for the velocity at first and second order in parametrized form (i.e., depending on $\fett{q}$), 
\begin{subequations}
\begin{align}
  & \theta_\alpha^{(1)}(\fett{x}(\fett{q})) =  \partial_D^{\rm L} \delta_\alpha^{(1)}(\fett{x}(\fett{q})) \,, \label{eqtheta1}\\
  & \theta_\alpha^{(2)}(\fett{x}(\fett{q})) =  \partial_D^{\rm L} \delta_\alpha^{(2)}(\fett{x}(\fett{q})) - \delta_\alpha^{(1)}(\fett{x}(\fett{q})) \, \theta_\alpha^{(1)}(\fett{x}(\fett{q}))  \label{eqtheta2}  \,.
\end{align}
\end{subequations}
Given the map and associated mass conservation law, we will solve these equations in the two Lagrangian approaches.
We also note that, alternatively, the Eulerian velocity divergence could be also determined by considering the convective time derivative of the Lagrangian displacement field, from which one subsequently needs to take the Eulerian divergence; for explicit instructions in the single-fluid case, see e.g.\ Section 6.3 of \cite{Rampf:2012xa}.

\paragraph*{Eulerian velocity in the Lagrangian approach of \S\ref{sec:altgrowingL2F}.}
In the most straightforward implementation of LPT, the component displacement reads
\be \label{eq:psi2lfuidsol-slavedrep}
  \fett{x}^{\alpha}(\fett{q},D) - \fett{q} = \fett{\xi}^\alpha(\fett{q},D) = D\,\fett{\xi}^{\rm m(1)}(\fett{q}) + D^2 \fett{\xi}^{\rm m(2)}(\fett{q}) \,,
\ee
where mass conservation reads in this case
\be \label{eq:masscomprep}
  \delta_\alpha(\fett{x}^\alpha(\fett{q},D)) = \frac{1+ \delta_\alpha^\ini}{\det [\delta_{ij} + \xi_{i,j}^\alpha]} - 1 \,.
\ee
For reference $\fett{\xi}^{\rm m(1)}(\fett{q})$ and $\fett{\xi}^{\rm m(2)}(\fett{q})$ coincide in the present approach with the single-fluid displacement, and are given in Eqs.\,\eqref{eq:sol2LPTsingle}. Expanding equation~\eqref{eq:masscomprep}
gives 
\begin{align} \label{eq:massLPTalpha1rep}
  &\delta_\alpha(\fett{x}^\alpha(\fett{q},D)) = \delta_\alpha^\ini(\fett{q}) + D \varphi_{,ll}^\ini(\fett{q}) + D \delta_\alpha^\ini \varphi_{,ll}^\ini \nonumber \\
  &\quad+ D^2 \left[ \frac 5 7 \varphi^\ini_{,ll} \varphi^\ini_{,mm} + \frac 2 7  \varphi^\ini_{,lm} \varphi^\ini_{,lm}  \right] 
     + O(3) \,.
\end{align}
Here it is important to note that the convective time derivative does not commute with the Eulerian derivative/position; since we have chosen to formulate the continuity equation by using the convective time derivative, pullback operations must be performed after the temporal derivatives are evaluated. Keeping this in mind, one obtains from the above expressions and from Eq.\,\eqref{eqtheta1} at first order
\be
  \theta_\alpha^{(1)}(\fett{x}(\fett{q}),D) =  \partial_D^{\rm L} \delta_\alpha^{(1)}(\fett{x}(\fett{q})) =  \varphi_{,ll}^\ini(\fett{q}) \,,
\ee
and subsequently at second order, using Eq.\,\eqref{eqtheta2},
\be
  \theta_\alpha^{(2)}(\fett{x}(\fett{q}),D) = D \left[ \frac 3 7 \varphi^\ini_{,ll} \varphi^\ini_{,mm} + \frac 4 7  \varphi^\ini_{,lm} \varphi^\ini_{,lm}  \right] \,.
\ee
What is left is evaluating all terms at the current position, which induces a second-order term stemming from $\theta_\alpha^{(1)}(\fett{x}(\fett{q}))= \varphi_{,ll}^\ini(\fett{q})$. The velocity divergence truncated to second order is then
\be \label{eq:thetaLPT2f}
  \theta_\alpha = - \varphi_{,ll}^\ini - D \left[ \frac 3 7 \varphi^\ini_{,ll} \varphi^\ini_{,mm}+ \varphi^\ini_{,l} \varphi^\ini_{,lmm} + \frac 4 7  \varphi^\ini_{,lm} \varphi^\ini_{,lm}  \right] \,,
\ee
which, as anticipated, agrees with the velocity divergence in the single-fluid case; cf.\,Eq.\,\eqref{eq:2SPT}.

\paragraph*{Eulerian velocity in the Lagrangian approach of \S\ref{sec:unpertcoord}.}
Also in this approach, we need to verify whether the fastest growing mode of velocity divergence for the components agrees with the one from the single fluid.
In the approach of \S\ref{sec:unpertcoord}, the map was found to be
\begin{align} \label{eq:mapfullapp}
  \begin{aligned}   \fett{x}_\alpha^{\rm full}  &= \fett{q} + D\,\fett{\zeta}^{\rm m(1)}(\fett{q}) + D^2 \fett{\zeta}^{\rm m(2)}(\fett{q}) 
     - D \zeta_{\alpha,l}^{\rm ini} \fett{\nabla} \varphi_{,l}^{\rm ini}  \\ 
  & \quad +  \fett{\nabla} \zeta_\alpha^{\rm ini} -  \nab^{-2} \fett{\nabla} ( \delta_\alpha^\ini \zeta_{\alpha,l}^{\rm ini})_{,l}   - \nab^{-2} \fett{\nabla} \mu_2(\zeta_\alpha^\ini) \,, 
   \end{aligned}
\end{align}
which is to be used with the mass conservation law
\be
  \delta_\alpha (\fett{x}^{\rm full}(\fett{q},D)) = \frac{1}{\det [x_{\alpha i,j}^{\rm full}(\fett{q},D) ]} - 1\,.
\ee 
Expanding this to second order we find
\begin{align}
   &\delta_\alpha (\fett{x}^{\rm full}(\fett{q},D)) = \delta_\alpha^\ini(\fett{q}) + D \varphi_{,ll}^\ini(\fett{q})   + D \zeta_{\alpha,l}^\ini \varphi_{,lmm}^\ini  - D \zeta_{\alpha,ll}^\ini \varphi_{,mm}^\ini \nonumber \\ 
&+ (\delta^\ini \zeta_{\alpha,l}^\ini)_{,l} + (\zeta_{\alpha,ll}^\ini)^2 
  + D^2 \left[ \frac 5 7 \varphi^\ini_{,ll} \varphi^\ini_{,mm} + \frac 2 7  \varphi^\ini_{,lm} \varphi^\ini_{,lm}  \right] .
\end{align}
Plugging this into the first-order perturbation equation~\eqref{eqtheta1} we find 
\be
 \theta_\alpha^{(1)}(\fett{x}^{\rm full}(\fett{q})) =   \varphi_{,ll}^\ini(\fett{q}) \,,
\ee
and from the second-order equation~\eqref{eqtheta2}
\be
 \theta_\alpha^{(2)}(\fett{x}^{\rm full}(\fett{q})) =  D \left[ \frac 3 7 \varphi^\ini_{,ll} \varphi^\ini_{,mm} + \frac 4 7  \varphi^\ini_{,lm} \varphi^\ini_{,lm} \! \right] 
    + \zeta_{\alpha,l}^\ini \varphi_{,lmm}^\ini  .
\ee
Finally, correcting for the position by taking the displacement~\eqref{eq:mapfullapp} into account, we obtain the truncated velocity divergence precisely as given in Eq.\,\eqref{eq:thetaLPT2f} and thus, also the present Lagrangian approach returns consistent results in terms of PT.

\subsection{Proof of zero relative displacement for growing modes}\label{app:xibc}

Here we show that the Lagrangian equations of motion~\eqref{eq:mainL2F} for the case of growing modes, where $\fett{\xi}^{\bc(1)}=0$ and $\delta_\rM^{\rm ini}=0$, predict vanishing relative displacements $\fett{\xi}^{\bc(n)}=\fett{\xi}^{\rb (n)}-\fett{\xi}^{\rc (n)}=0$ at all orders in LPT. We will prove this by induction, showing that if $\xi^{\bc (k)}=0$ for all $k<n$ then $\xi^{\bc (n)}=0$. 

We begin by computing the difference of the evolution equations for the component displacements~\eqref{eq:LEOM1}
and replacing $x^\alpha_{i,j}=\delta_{ij}+\xi^\alpha_{i,j}$ on the LHS, which leads to
\begin{align} 
 &2 \RT \xi^{\bc (n)}_{i,i}+2 \varepsilon_{ikl} \varepsilon_{jkn}   \left( \xi_{l,n}^{\rm b} \RT \xi_{i,j}^{\rm b} - \xi_{l,n}^{\rm c} \RT \xi_{i,j}^{\rm c}\right)^{(n)} \notag \\
 &\quad  + \varepsilon_{ikl} \varepsilon_{jmn} \left( \xi_{k,m}^{\rm b}  \xi_{l,n}^{\rm b} \RT \xi_{i,j}^{\rm b} - \xi_{k,m}^{\rm c}  \xi_{l,n}^{\rm c} \RT \xi_{i,j}^{\rm c} \right)^{(n)} \notag\\
 &= - \frac{3g}{D} \left[J^\rb(\nab_x^2 \varphi)_\rb- J^\rc(\nab_x^2 \varphi)_\rc\right]^{(n)}\,.
\end{align}
Evidently, the round bracketed term on the LHS contains terms quadratic and cubic in the displacements. Since their orders need to sum to $n$, all displacements involved will be evaluated at an order $k<n$, where 
it has already been established that $\fett{\xi}^{\rb (k)} = \fett{\xi}^{\rc (k)}$  for $k<n$ and thus, these terms cancel out.
This leads to
\begin{align} 
\label{eq:LPTlhs}
 \RT \xi^{\bc (n)}_{i,i} &=  -\frac{3g}{2D} \left[ J^\rb(\nab_x^2 \varphi)_\rb- J^\rc(\nab_x^2 \varphi)_\rc \right]^{(n)}  \nonumber \\
  &=  -\frac{3g}{2D} \Bigg[ \fb (1+\delta_\rb^\ini) + \fc (1+\delta_\rc^\ini) \frac{J^\rb(\fett{q})}{J^\rc(\tilde{\fett{q}}_1)} - J^\rb  \nonumber \\
         &\quad\,  - \fb (1+\delta_\rb^\ini) \frac{J^\rc(\fett{q})}{J^\rb(\tilde{\fett{q}}_2)} 
       -\fc (1+\delta_\rc^\ini)  + J^\rc  \Bigg]^{(n)},
\end{align}
where $\tilde{\fett{q}}_1 = \fett{x}_\rc^{-1} \circ \,\fett{x}^{\rm b}(\fett{q})$ and $\tilde{\fett{q}}_2 =  \fett{x}_{\rm b}^{-1} \circ\,\fett{x}^{\rm c}(\fett{q})$. To arrive at the last equality, we have used the Poisson equations~\eqref{eq:deltamLag} for the two fluids.
To proceed note that the $n$th-order approximations for the Jacobians can be exactly written as
\begin{align} \label{eq:Jalpha}
  \left[ J^\alpha(\fett{q})-1 \right]^{(n)} &=  \xi_{l,l}^{\alpha (n)}(\fett{q}) + \delta F^{\alpha (n)}(\fett{q}) \,,
\end{align}
and specifically
\be \label{eq:Jctilde}
 \begin{aligned}
   \Big[ J^\rc(\tilde{\fett{q}}_1)-1 \Big]^{(n)} &= \xi_{l,l}^{\rc (n)}(\fett{q}) + \delta F^{\rc (n)}(\tilde{\fett{q}}_1) \,, \\
    \Big[ J^\rb(\tilde{\fett{q}}_2)-1 \Big]^{(n)} &= \xi_{l,l}^{\rb (n)}(\fett{q}) + \delta F^{\rb (n)}(\tilde{\fett{q}}_2) \,,
 \end{aligned}
\ee
where $\delta F^{\alpha(n)}(\fett{q}) = [\mu_2^\alpha + \ldots]^{(n)}$, and for the 
specific cases 
$\delta F^{\rc (n)}(\tilde {\fett{q}}_1) = [\mu_2^\rc + \xi_{l,lm}^\rc \xi_{m}^{\bc} + \ldots]^{(n)}$
and 
$\delta F^{\rb (n)}(\tilde {\fett{q}}_2) = [\mu_2^\rb + \xi_{l,lm}^\rb \xi_{m}^{\bc} + \ldots]^{(n)}$
where we have used Eqs.\,\eqref{eqJalp}--\eqref{eqJc}.
 Thus, the functions $\delta F^{\alpha (n)}$ are {\it combinations of quadratic and cubic lower-order perturbations}; but since we have established at lower orders that $\fett{\xi}^{\rb (k)} = \fett{\xi}^{\rc (k)}$  for $k<n$, it follows that all the occurring $\delta F^{\alpha (n)}$ in Eq.\,\eqref{eq:Jalpha}--\eqref{eq:Jctilde} are actually identical.
From this it follows that 
\be
  \left[\frac{J^\rb(\fett{q})}{J^\rc(\tilde{\fett{q}}_1)} - 1\right]^{(n)}\!\! =  \xi_{l,l}^{\rb (n)} - \xi_{l,l}^{\rc (n)} , \quad  \left[\frac{J^\rc(\fett{q})}{J^\rb(\tilde{\fett{q}}_2)} - 1\right]^{(n)} \!\!=  \xi_{l,l}^{\rc (n)} - \xi_{l,l}^{\rb (n)} .
\ee
Using this in Eq.\,\eqref{eq:LPTlhs} we then establish
\be
  \RT \xi^{\bc (n)}_{i,i} = 0\,,
\ee
and thus, $\xi^{\bc (n)}_{i,i}=0$ at all successive orders if $\xi^{\rm bc (k)}_{i,i}=0$ for $k<n$, which concludes the proof.

\section{Contact Hamiltonian in \texorpdfstring{$\fett{\Lambda}$}{L}CDM}\label{app:HJ}

Here we derive the so-called contact Hamiltonian~\eqref{eg:contactH}, as well as provide some related tools. Many of the results presented here follow the methodology of~\cite{BRAVETTI201717}, applied to the cosmological problem; for more mathematical details we refer to their work and references therein.

 Contact Hamiltonians employ contact geometry, a concept introduced through~\cite{Arnold1989}. Contact Hamiltonians  are particularly well suited for dissipative systems and thus, due to the continuous energy extraction -- thanks to the comoving expansion of the Universe, are ideally suited for our purpose. One of the central ideas of contact Hamiltonians is to extend the symplectic phase-space ($3+3$ dimensions) by an extra dimension. Historically, such extended phase-spaces incorporated the time as the additional dimension, however within the contact formalism one chooses instead a non-trivial dynamical variable. It turns out that, up to an additive constant, this variable is in fact the action
\be
   {\cal S}(\fett{x}, D) = \int {\cal L}(\fett{x}, \dot{\fett{x}}, D) \, \dd D = \int \left(\frac{\partial \cal L}{\partial \dot{\fett x}} \cdot \dot{\fett{x}}  - H\right) \, \dd D 
\ee
 of the system, where one can rewrite the first term using $\dot{\fett x}\, \dd D = \dd \fett{x}$.

\cite{BRAVETTI201717} derived the contact transformations, which are essentially the counterpart of canonical transformations in standard Hamiltonian mechanics; for an arbitrary transformation of coordinates $(x^i, p_i, S) \to (\tilde x^i, \tilde p_i, \tilde S)$ they are \citep[cf.\ Eqs.\,(60-62) of][]{BRAVETTI201717}
\begin{subequations} \label{eq:contacttrafos}
\begin{align}
  \frac{\partial \tilde {\cal S}}{\partial {\cal S}} - \tilde p_l \frac{\partial \tilde x^l}{\partial {\cal S}} &= {\mathfrak f} \,, \\
  \frac{\partial \tilde {\cal S}}{\partial x^i} - \tilde p_l \frac{\partial \tilde x^l}{\partial x^i} &= - {\mathfrak f}\, p_i \,, \\
  \frac{\partial \tilde {\cal S}}{\partial p_i}  - \tilde p_l \frac{\partial \tilde x^l}{\partial p_i} &= 0\,,
\end{align}
while the contact transformation of the Hamiltonian ${\cal H} \to \tilde {\cal H}$ is \citep[cf.\ Eqs.\,(82) of][]{BRAVETTI201717}
\be \label{eq:trafocontactH}
  \frac{\partial \tilde {\cal S}}{\partial D} - \tilde p_l \frac{\partial \tilde x^l}{\partial D} + \tilde {\cal H} = {\mathfrak f}\, {\cal H} \,,
\ee
\end{subequations}
where ${\mathfrak f}$ is a (time-dependent) parameter that needs to be determined.

Now we are equipped to determine the contact Hamiltonian in $\Lambda$CDM. 
For this let us begin with the Hamiltonian in $D$-time (Eq.\,\ref{eg:contactH}) formulated in the extended phase-space, i.e., including the functional dependence of ${\cal S}$,
\be \label{eq:Hextended}
  {\cal H}(\fett{x},\fett{p}, {\cal S}, D) =  \frac{\fett{p}^2}{2 m a^2 (\partial_t D)} + \frac{3D}{2a (\partial_t D)} \varphi(\fett{x}) \,,
\ee
and consider the time-dependent contact transformation
\be
  \left( \fett{x},\fett{p}, {\cal S}, D \right) \to \left( \tilde {\fett{x}}, \tilde{\fett{p}}, \tilde {\cal S}, D \right) \,.
\ee
It is easily checked that the following transformation $\fett{x}=\tilde {\fett{x}}$,  $\fett{p} =  a^2 (\partial_t D) \fett{\tilde p}$, and ${\cal S} = a^2 (\partial_t D) \tilde {\cal S}$ is in accordance with~Eqs.\,\eqref{eq:contacttrafos} and thus, is indeed a contact transformation.This particular transformation is simple in the sense that the first three equations from \eqref{eq:contacttrafos} just define $\mathfrak f=  [a^2 (\partial_t D)]^{-1}$ and the contact-transformed Hamiltonian is given by $\tilde{\cal H}=\mathfrak f{\cal H} - \partial \tilde {\cal S}/\partial D$ following Eq.\,\eqref{eq:trafocontactH}. The latter term can be simplified using the equation~\eqref{eq:evoL} for the linear growth rate  which is equivalently $\partial_t (a^2 \partial_t D) = 3D/(2a)$ \citep[c.f.][]{Brenier:2003xs}, where $g=(D/\partial_t D)^2 a^{-3}$. Hence, the Hamiltonian~\eqref{eq:Hextended} turns into the contact transformed Hamiltonian 
\be \label{eq:contactHrep}
  \tilde{\cal H}(\tilde{\fett{x}}, \tilde{\fett{p}}, \tilde{\cal S},D) = \frac{{\tilde{\fett{p}}}^2}{2m} + \frac{3g}{2D} \left[ \tilde{\cal S} + \varphi(\tilde{\fett{x}}) \right] \,,
\ee
which, supplemented with the replacement $\fett{\tilde p} \to \fett{v}$ and $m=1$, agrees with the one in the main text, where we remove all tildes to avoid unnecessary cluttering.

Finally, for reasons of completeness let us report the equations of motions for the contact Hamiltonian, governed not by the usual Hamilton equations, but instead by 
\begin{subequations}
\begin{align}
   \frac{\dd x^i}{\dd D}  &= \frac{\partial {\cal H}}{\partial p_i} \,, \\
  \frac{\dd p_i}{\dd D}  &= -\frac{\partial {\cal H}}{\partial x^i} - p_i \frac{\partial {\cal H}}{\partial S} \,, \\
   \frac{\dd  {\cal S}}{\dd D}  &= p_l \frac{\partial {\cal H}}{\partial p_l} - {\cal H} 
\end{align}
\end{subequations}
\citep[cf.\ Eqs.\,(37-39) of][]{BRAVETTI201717},
which in the case of~\eqref{eq:contactHrep}, for $\fett{p} \to \fett{v}$ and $m=1$, respectively lead to Eqs.\,\eqref{eq:contactEOMs} in the main text, which we repeat here for convenience,
\begin{subequations}
\begin{align}
   \frac{\dd \fett{x}}{\dd D}  &= \fett{v}  \,, \\
   \frac{\dd \fett{v}}{\dd D}  &= - \frac{3g}{2D} \left( \fett{v} + \nab \varphi \right) \,, \\
   \frac{\dd {\cal S}}{\dd D}  &= \frac{\fett{v}^2}{2} - \frac{3g}{2D} \left(  \varphi  +  {\cal S}  \right) \,. \label{eq:actionD}
\end{align}
\end{subequations}
These equations, {\it which remain regular at $D \to 0\,$ for slaved boundary conditions}~\eqref{eq:slaving}, may be solved perturbatively in a similar fashion as outlined in the main text, which then lead to a contact Hamiltonian-style perturbation theory -- an avenue that so far has not been reported in the cosmological literature. Furthermore, from~\eqref{eq:actionD} one may determine the action  perturbatively.

Such a contact-Hamiltonian perturbative expansion would be similar, yet fairly distinct to the approaches of \cite{Bartelmann:2014fma,Floerchinger:2016hja,McDonald:2018,2019JCAP...04..001L,2019JCAP...05..017G}, who perform expansions around (parts of) the Hamiltonian (or action), which is however incompatible with slaving boundary conditions that we employ in this paper. Indeed we have shown that not the Hamiltonian but the contact transformed Hamiltonian  allows for growing-mode solutions, in accordance with slaving.

\hfill

\bsp	
\label{lastpage}
\end{document}